\documentstyle[12pt,epsfig]{article}
\newlength{\aivwidth}   
\newlength{\tmpwidth}   

\newcommand{\lsim}
{\;\raisebox{-.3em}{$\stackrel{\displaystyle <}{\sim}$}\;}
\newcommand{\gsim}
{\;\raisebox{-.3em}{$\stackrel{\displaystyle >}{\sim}$}\;}

\def\ew{\alpha_W}
\def\ewp{\alpha_{W\Phi}}
\def\ebp{\alpha_{B\Phi}}

\begin{document}

\setcounter{page}{1}
\thispagestyle{empty}

\large
\begin{flushright}
hep-ph/9706406\\
BI-TP 97/16\\
PM/97-09\\
June 1997
\end{flushright}
\vspace*{8mm}
\large
\begin{center}
{\bf Gauge-Boson Pair Production at the LHC:\\
Anomalous Couplings and Vector-Boson Scattering\footnote{
Supported by the EC network contract CHRX-CT94-0579 and the
BMBF, Bonn, Germany}\\}
\vspace*{10mm}
\large I. Kuss$^{\scriptstyle 1}$, E. Nuss$^{\scriptstyle 2}$\\
\normalsize
{\em $^1$Fakult\"at f\"ur Physik, Universit\"at Bielefeld,}\\
{\em Postfach 10 01 31, 33501 Bielefeld, Germany}\\
{\em $^2$Physique Math\'{e}matique et Th\'{e}orique, CNRS-URA 768,}\\
{\em Universit\'{e} de Montpellier II, F-34095 Montpellier Cedex 5, France}
\end{center}
\normalsize
\vspace*{8mm}

\begin{abstract}
We compare vector boson fusion and quark antiquark annihilation
production of vector boson pairs at the LHC and include the effects of
anomalous couplings. Results are given for confidence intervals for
anomalous couplings at the LHC assuming that measurements will be in
agreement with the standard model. We consider all couplings of the
general triple vector boson vertex and their correlations. 
In addition we consider a gauge invariant dimension-six
extension of the standard model. Analytical results for the
cross sections for quark antiquark annihilation and vector boson
fusion with anomalous couplings are given.
\end{abstract}

\section{Introduction}
In this note we study vector boson pair production with possible
anomalous couplings in proton proton collisions at the LHC.
The motivation to study these
processes has been twofold:
\begin{enumerate}
\item
If the electroweak symmetry breaking is not realized by a
light Higgs boson, the symmetry breaking will manifest itself
by some strong interactions among longitudinally polarized gauge bosons
\cite{strong_0,dimo}. In general, the amplitudes for 
longitudinal vector boson scattering
are then very large at high energies.
Several models to describe the strongly coupled symmetry
breaking, in particular the standard model with a heavy Higgs boson
and technicolor inspired models, have been discussed \cite{philips,dobado1,
dobado,goldplate}.   
If an amplitude has been calculated within a
specific model,
a method to connect this amplitude to parton parton scattering 
processes has to be employed. The conventional
method \cite{chaga,dkr,strong_2,falk} was to use the
effective vector boson approximation (EVBA) \cite{evba}.
The EVBA was originally used only for longitudinally polarized vector
bosons. It was however also applied to 
all intermediate helicity states \cite{dobado} despite of the known
problems with the EVBA for the
transverse helicities \cite{godbole}. 

Avoiding the use of the EVBA, the quark-quark scattering processes
$q_1q_2\to V_3V_4 q_1'q_2'$ were calculated exactly in lowest order
of perturbation theory \cite{philips,goldplate}. 
In addition to the vector boson scattering diagrams,
diagrams of bremsstrahlung type have to be evaluated in the exact
calculation. 

\item On the other hand one may assume that the symmetry breaking is
realized by a light Higgs 
boson. In this case the dominating processes for vector boson pair
production are those of direct quark antiquark annihilation, 
also called Drell-Yan
processes. The rates for these processes are sensitive to the values of
the couplings of the electroweak vector bosons among each other
\cite{kuijf}.
Drell-Yan production with anomalous (=non-standard) couplings has been
studied in
\cite{aihara}-\cite{berger}.
${O}(\alpha_s)$ corrections have been taken into account
in \cite{alpha_s}-\cite{bho_ww}. 
The vector boson scattering processes were not considered in these
works.
The common argument to omit these processes was
that they
are ${O}(\alpha^4)$ and therefore suppressed with
respect to the Drell-Yan processes. 
However, a particular case in which these processes can give a
significant contribution
is near a Higgs boson resonance.
In the study \cite{Higgs1} of the signal of a resonant Higgs boson
both the Drell-Yan processes, including the ${O}(\alpha_s)$
corrections \cite{Higgs2}, 
and the exact matrix element for $q_1q_2\to V_3V_4q_1'q_2'$ 
were taken into account.
Also in \cite{2782},
the processes $q_1q_2\to V_3V_4q_1'q_2'$
were included.
These calculations however were only for standard vector boson self
couplings and the rates for the two different production mechanisms
have not been explicitly compared.
\end{enumerate}
In summary, in the strongly interacting scenario particular attention
was paid to the vector boson scattering processes while the analyses of
vector boson self couplings only considered the Drell-Yan processes.

Later on,
the effects of various $SU(2)_L\times U(1)_Y$ gauge invariant
effective interaction terms among the electroweak
vector bosons were investigated and the vector boson
scattering processes were considered \cite{renard,amps} together with
the Drell-Yan processes. 
It was found that the Drell-Yan contribution and the one of vector boson
scattering were of comparable magnitude.
However, as in \cite{dobado}, 
the vector boson scattering processes were calculated using the EVBA for
all intermediate boson helicities.

Recently \cite{kussspie,kuss}
we showed that an improved version of the
EVBA can increase the reliability of EVBA calculations.
In particular, the improved EVBA could well reproduce the
result of a
complete perturbative calculation for a process which is dominated
by the transverse intermediate helicities.

In this article we carry out a comparison of Drell-Yan production and
vector boson scattering using the improved EVBA and
including the influence of anomalous couplings.
This work is thus a supplement to the existing analyses 
\cite{aihara,barger,baur,bho_ww}
in which the Drell-Yan processes have been considered
in more detail ($O(\alpha_s)$ corrections were included and more refined
kinematical cuts were applied), but vector boson scattering was not discussed.
We will study the general
parametrization \cite{kmss2},\cite{gaemers}-\cite{laymou} 
of the triple gauge
boson vertices in terms of seven parameters,
allowing for $C$- and $P$-violation.
In addition, we will study an $SU(2)_L\times U(1)_Y$
gauge invariant dimension-six extension of
the standard model. Our work
extends the works \cite{renard,amps} in that all three
$C$- and $P$-invariant gauge invariant
dimension-six operators \cite{rujula}-\cite{gids} 
which affect the vector boson self-interactions are discussed.
We note that the three $C$- and $P$-invariant trilinear couplings which
potentially contribute to the experimentally relevant \cite{kuijf}
process of $W^{\pm}Z$ Drell-Yan production can be equivalently expressed
in terms of the parameters of the three-parameter gauge invariant model.
The same is true for the two $C$- and $P$-invariant couplings which potentially
contribute to the similarly relevant process of $W^{\pm}\gamma$
production.

In Section \ref{compsec} we compare vector boson fusion and Drell Yan
production in the three-parameter gauge invariant model.
In Section \ref{fits}, we present parameter 
fits for the anomalous couplings which can be obtained from future
LHC measurements assuming that standard model predictions will
actually be measured. We discuss the full set of anomalous couplings
and also give the unitarity limits for the set of couplings which we 
use. We also consider again the three-parameter gauge
invariant model.
In Appendix \ref{appa} we give analytical formulas for the cross
sections for $q\bar{q}'$ annihilation into $W^\pm Z,W^\pm\gamma$ and
$W^+W^-$ pairs in terms of the seven anomalous couplings.
In Appendix \ref{appb} we give formulas for vector boson
scattering cross sections for the gauge invariant model.

\section{Comparison of Vector-Boson Fusion and Drell-Yan Production
\label{compsec}}
\begin{figure}
\begin{center}
\epsfig{file=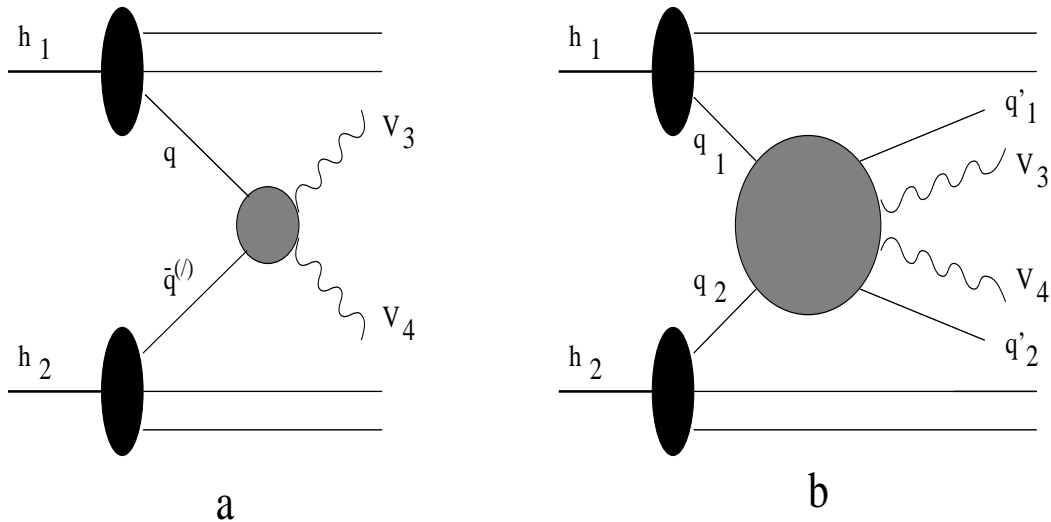,width=14cm,height=7cm}
\end{center}
\caption{Diagrammatic representations for the production of a
vector boson pair $V_3V_4$ in the collision of two hadrons $h_1h_2$.
a: via the quark antiquark annihilation mechanism.
b: via the ${O}(\alpha^4)$ parton reaction 
   $q_1q_2\to V_3V_4q'_1q'_2$.}
\label{comp}
\end{figure}
To illustrate our results we
calculate the invariant mass distributions of the cross sections
for vector boson pair production
at the LHC ($pp$ collisions at $\sqrt{s_{pp}}=14$ TeV).
We compute both the contribution from Drell-Yan production and from
the ${O}(\alpha^4)$ parton reaction $q_1q_2\to V_3V_4q_1'q_2'$
which contains vector boson scattering. The two contributions are shown
diagrammatically in Figure \ref{comp}. 
Both contributions are calculated in the Born approximation and
we use the improved EVBA \cite{kussspie,kuss} to
calculate the latter contribution.
We discuss all possible pairs of produced electroweak vector-bosons, $W^{\pm}Z,
W^{\pm}\gamma,W^+W^-,ZZ,W^{\pm}W^{\pm},Z\gamma$ and $\gamma\gamma$.
We first present the results for the standard model and then
for non-zero anomalous couplings.

\subsection{Calculational Procedure}
\subsubsection{Drell-Yan Production}
\begin{figure}
\begin{center}
\epsfig{file=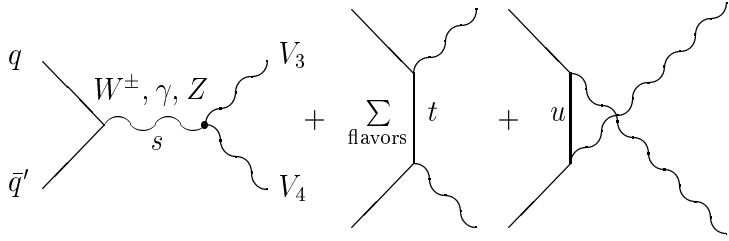,width=10cm,height=3.5cm}
\end{center}
\caption{The generic Feynman diagrams for 
a process $q\bar{q}'\to V_3V_4$ in lowest order of perturbation theory.}
\label{qqdia}
\end{figure}

In the usual quark-parton description, the lowest order contribution
comes from the Drell-Yan processes shown in 
Figure \ref{comp} (a). 
Three generic Feynman diagrams can contribute to any
of these processes in lowest order (Fig. \ref{qqdia}).
They correspond to the exchange of vector boson(s) in the
$s$-channel and the exchange of fermions in the
$t$- and in the $u$-channel.
Only the vector-boson exchange diagrams receive a contribution from the
vector boson self-couplings.
The standard model
differential cross sections for $q\bar{q}'\to V_3V_4$ have been first
given in \cite{brown}. The results for arbitrary $\ew$
can be found in \cite{renard}.
For arbitrary vector boson
self couplings, demanding only Lorentz-invariance, 
the differential cross sections as well as the expressions for the
helicity amplitudes
have been recently given for all processes in analytical form in
\cite{nuss}. 
We choose to repeat the formulas for the differential cross sections
in Appendix \ref{appa}
in a form in which the high energy behavior is immediately transparent.
We note that the ${O}(\alpha_s)$-corrections
to the lowest order cross-sections can be huge. 
For $W^{\pm}Z$ production \cite{baur} they can reach up to 70\% of
the lowest order contribution  and for $W^{\pm}\gamma$ production 
\cite{berger} they can be even larger. 
Only the Born cross section will be considered here. 

The formula for the invariant mass distribution of the cross section
for $V_3V_4$-pair production
via $q\bar{q}'$-annihilation in the collision of two hadrons $h_1h_2$
is given by
\begin{eqnarray}
&&\frac{d\sigma}{dM_{V_3V_4}}
(h_1 h_2\to q\bar{q}'\to V_3 V_4,s_{hh})|_{\mathrm{cut}}\cr 
&=&\frac{2M_{V_3V_4}}{s_{hh}}\int\limits_{-y_{\mathrm{max}}}^{
y_{\mathrm{max}}}dy\sum_{(q\bar{q}')}
\left[f_{q}^{h_1}(\sqrt{x}e^{y},Q_1^2)f_{\bar{q}'}^{h_2}(\sqrt{x}e^{-y},Q_2^2)
\;\;+\;\;h_1\leftrightarrow h_2\;\right]\cr
&&\quad\quad\quad
\times\int\limits_{z_{\mathrm{min}}(y)}^{
z_{\mathrm{max}}(y)}d\cos\theta\frac{d\sigma}{d\cos\theta}
(q\bar{q}'\to V_3 V_4).
\label{sigqq}
\end{eqnarray}
This formula is valid if either no cuts or a rapidity or a pseudorapidity
cut on both produced vector bosons is applied. 
A pseudorapidity cut, in contrast to a rapidity cut, always excludes
events near the hadron beam direction.
In (\ref{sigqq}), $\sqrt{s_{hh}}$ and
$M_{V_3V_4}$ are the invariant masses of the hadron pair and the
vector boson pair, respectively, 
$y$ is the rapidity of the $q\bar{q}'$-pair in the
$h_1h_2$ c.m.s and 
$x\equiv M_{V_3V_4}^2/s_{hh}$.
The quantities $f_q^{h_i}$ denote the
parton distributions in the hadrons and the quantities
$Q_i^2$ are the factorization
scales. $\theta$ is the angle between the
quark $q$ and the vector-boson $V_3$ in the center-of-mass 
system of the quarks.
Applying no cuts, the limits of integration in (\ref{sigqq}) are 
$y_{\mathrm{max}}=\frac{1}{2}\ln(1/x)$
and $z_{\mathrm{min}}(y)=z_{\mathrm{max}}(y)=1$. 
We choose here to apply a pseudorapidity cut $\eta$
on the produced vector bosons. This cut is equivalent to a minimum required
angle $\vartheta_{\mathrm{min}}$ between the direction of momentum of any
of the produced 
vector-bosons
and the hadron beam direction. The cut $\eta$ is related to 
$\vartheta_{\mathrm{min}}$ by $\tanh(\eta)\equiv\cos\vartheta_{\mathrm{min}}$.
The integration limits with an $\eta$-cut in the $h_1h_2$ c.m.s
are given by
\begin{eqnarray}
y_{\mathrm{max}}&=&\min\left[\frac{1}{2}\ln\left(\frac{1}{x}\right),
\tanh^{-1}\left(\sqrt{\frac{1}{1+(\min(E_3^2,E_4^2)/q^2)\tan^2
\vartheta_{\mathrm{min}}}}\right),\right.\cr
&&\left.\quad\quad\quad\quad\quad\quad\quad\!
\tanh^{-1}\left(\sqrt{\frac{1}{1+(\max(M_3^2,M_4^2)/q^2)\sin^2
\vartheta_{\mathrm{min}}}}\right)\right],\cr
z_{\mathrm{min}\atop\mathrm{max}}(y)&=&
\frac{1}{q(1+t^2\gamma^2)}
{\max\atop\min}\left[-t^2\gamma^2\beta E_3
\mp\sqrt{q^2(1+t^2\gamma^2)-t^2\gamma^2\beta^2E_3^2}\right.,\cr
&&\quad\quad\quad\quad\quad\quad\quad\quad\;\,
\left. t^2\gamma^2\beta E_4
\mp\sqrt{q^2(1+t^2\gamma^2)-t^2\gamma^2\beta^2E_4^2}\right],
\label{zlimits}\end{eqnarray}
and one has to require that $z_{\mathrm{min}}<z_{\mathrm{max}}$.
The upper sign of $\mp$ in (\ref{zlimits}) is for
$z_{\mathrm{min}}$, the lower sign for $z_{\mathrm{max}}$.
In (\ref{zlimits}), 
the quantity $\beta\equiv\tanh(y)$ is the  boost-parameter for
a transformation from the $(h_1h_2)$ c.m.s into the $(V_3V_4)$
(=$(q\bar{q}')$) c.m.s. and 
$t^2\equiv\tan^2\vartheta_{\mathrm{min}}$. Further we defined
$\gamma^2\equiv1/(1-\beta^2)$.
The quantities $E_3\equiv\sqrt{q^2+M_3^2}$ and $E_4\equiv\sqrt{q^2+M_4^2}$
are the energies of $V_3$ and $V_4$ in the $(q\bar q')$ cms, 
while $M_3$ and $M_4$ are their masses.
$q$ is the magnitude of the three-momentum of $V_3$ or $V_4$ in the
$(V_3V_4)$ cms-system.
The last argument of the $\min$-function which defines $y_{\mathrm{max}}$
in (\ref{zlimits}) only plays a role near the threshold.
In deriving (\ref{zlimits}) we assumed that the quarks have no transverse
momentum with respect to the hadrons, but no other kinematical approximations
were made.

For large energies of the produced
vector-bosons,
$q^2\gg \max(M_3^2,M_4^2)$, the limits of integration 
(\ref{zlimits}) take on the simplified forms
\begin{eqnarray}
y_{\mathrm{max}}&\simeq&\min\left[\frac{1}{2}\ln\left(1/x\right),
\eta\right],\cr
z_{\mathrm{max}}(y)&\simeq&-z_{\mathrm{min}}(y)\;\simeq\;\tanh(\eta-|y|).
\label{etacut_limits}\end{eqnarray}
In this limit, the $\eta$-cut is identical to a rapidity-cut $Y$ of the same
magnitude.
We choose a cut of the magnitude $\eta=1.5$, 
corresponding to a minimum angle of
$\theta_{\mathrm{min}}=25^0$.
For the relevant process $pp\to W^{\pm}Z+X$
the highest sensitivity to anomalous couplings is achieved with
a cut of about this magnitude \cite{diss}.

\subsubsection{Vector Boson Fusion}
The ${O}(\alpha^4)$ partonic reaction which is shown in 
Figure \ref{comp} (b) contains 
the vector-boson scattering processes $V_1V_2\to V_3V_4$ as subprocesses.
Three types of Feynman diagrams contribute to a generic process
$V_1V_2\to V_3V_4$. They correspond to vector boson exchange, a direct
interaction among the four vector bosons and Higgs boson exchange.
Using the Feynman rules for the GIDS model given in \cite{dipl} 
we wrote the amplitudes
as functions of the scalar products of the external momenta and of
polarization vectors. We evaluated them numerically
without making further approximations.
In Appendix \ref{appb} we give analytical expressions for the cross 
sections for 
$W^{\pm}Z\to W^{\pm}Z$, $W^{\pm}\gamma\to W^{\pm}Z$ 
and $W^\pm\gamma\to W^\pm\gamma$ in a high energy approximation.
Expressions for the amplitudes of these and other
vector boson scattering processes 
can be found in \cite{amps,dipl,uni}.

We calculate the invariant mass
distribution of the cross-section for $h_1h_2\to V_1V_2\to V_3V_4$
in the improved EVBA according to
\cite{kuss}. The formulas which have been given there apply
if a rapidity cut is used. 
The corresponding expressions
for a  pseudorapidity cut are obtained by replacing $z_{\mathrm{min}}(y),
z_{\mathrm{max}}(y)$ and
$y_{\mathrm{max}}$ in \cite{kuss} by the expressions (\ref{zlimits}). 
We use the exact vector boson pair luminosities of \cite{kuss}
if $V_1V_2$ consists of two massive vector bosons.
If a photon is involved, the Approximation 2 of \cite{kuss}
with the photon distribution function of \cite{gsv} is used. 

\subsection{Results in the Standard Model}
\begin{figure}
\begin{center}
\epsfig{file=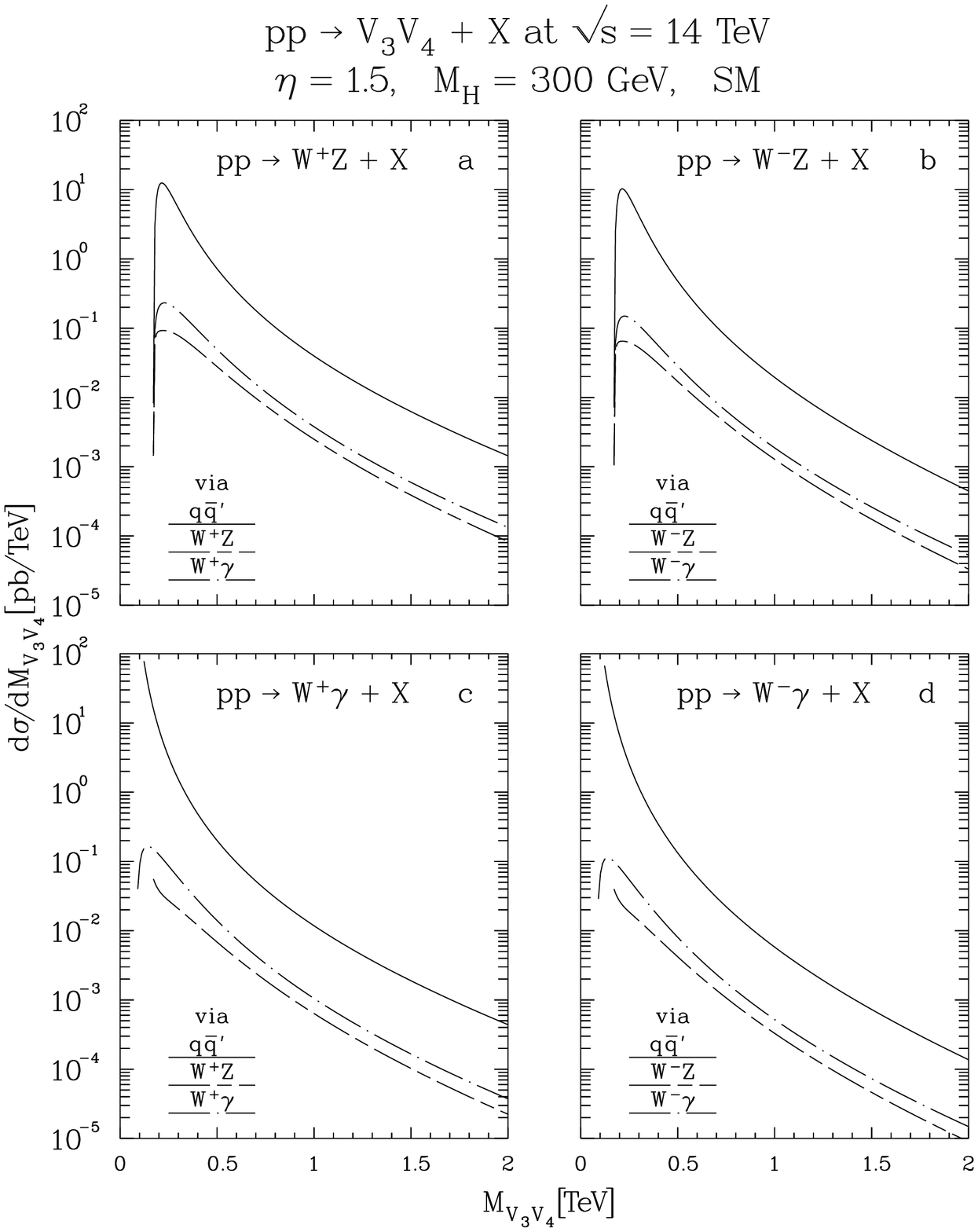,width=16cm,height=20cm}
\end{center}
\caption{The cross sections for $W^+Z,W^-Z,W^+\gamma$ and $W^-\gamma$
production as functions of the invariant mass $M_{V_3V_4}$ of the produced
vector boson pair for $pp$-collisions
at $\protect\sqrt{s}=14$ TeV. Separately
shown are the contributions from $q\bar q'$-annihilation and from vector-boson
fusion processes. A rapidity cut of $\eta=1.5$ has been applied.}
\label{fig1}
\end{figure}

\begin{figure}
\begin{center}
\epsfig{file=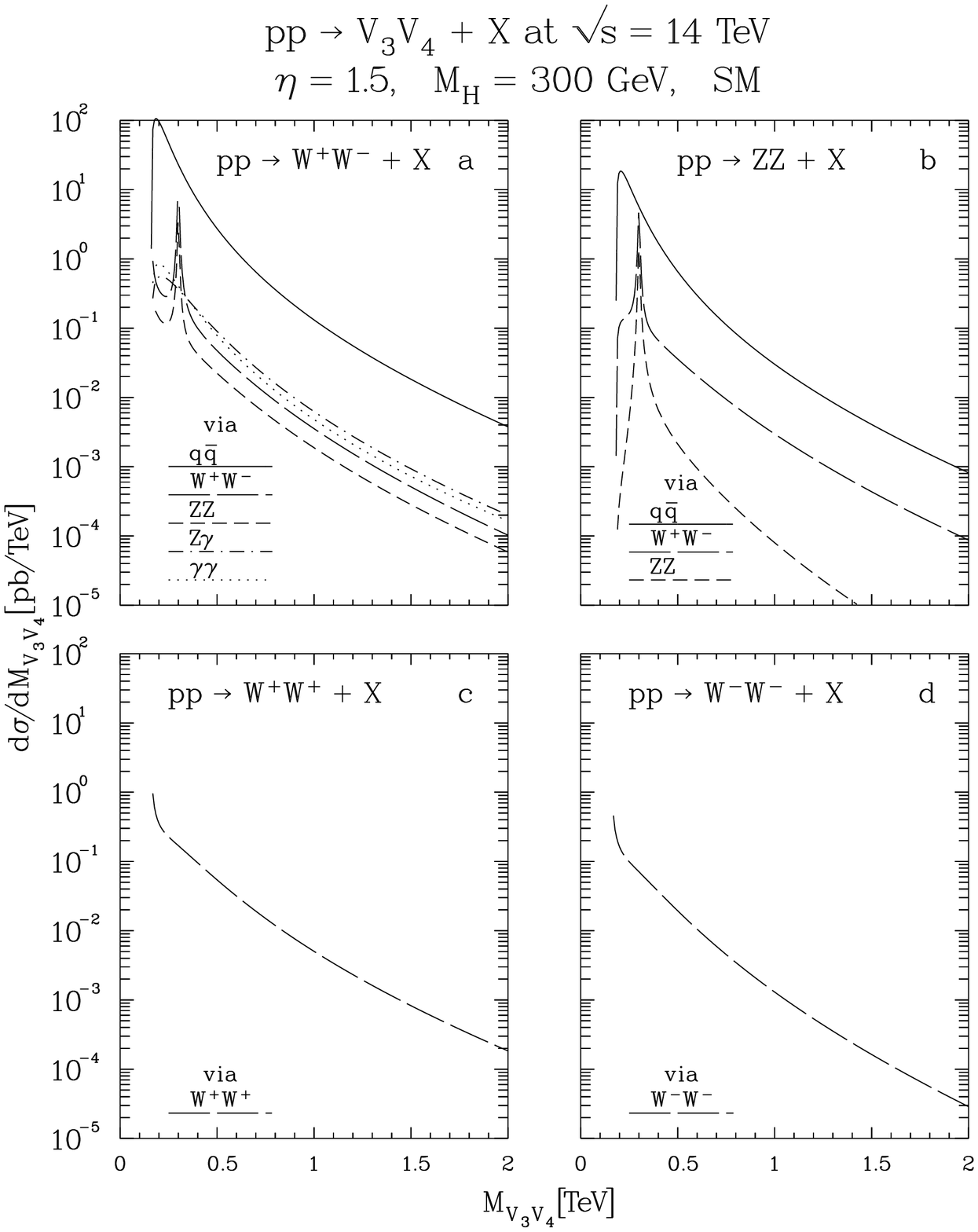,width=16cm,height=20cm}
\end{center}
\caption{The cross sections for $W^+W^-,ZZ,W^+W^+$ and $W^-W^-$
production as functions of the invariant mass $M_{V_3V_4}$ of the produced
vector boson pair for $pp$-collisions
at $\protect\sqrt{s}=14$ TeV. Separately
shown are the contributions from $q\bar q'$-annihilation and from vector-boson
fusion processes. A pseudorapidity cut of $\eta=1.5$ has been applied.}
\label{fig2}
\end{figure}

\begin{figure}
\begin{center}
\epsfig{file=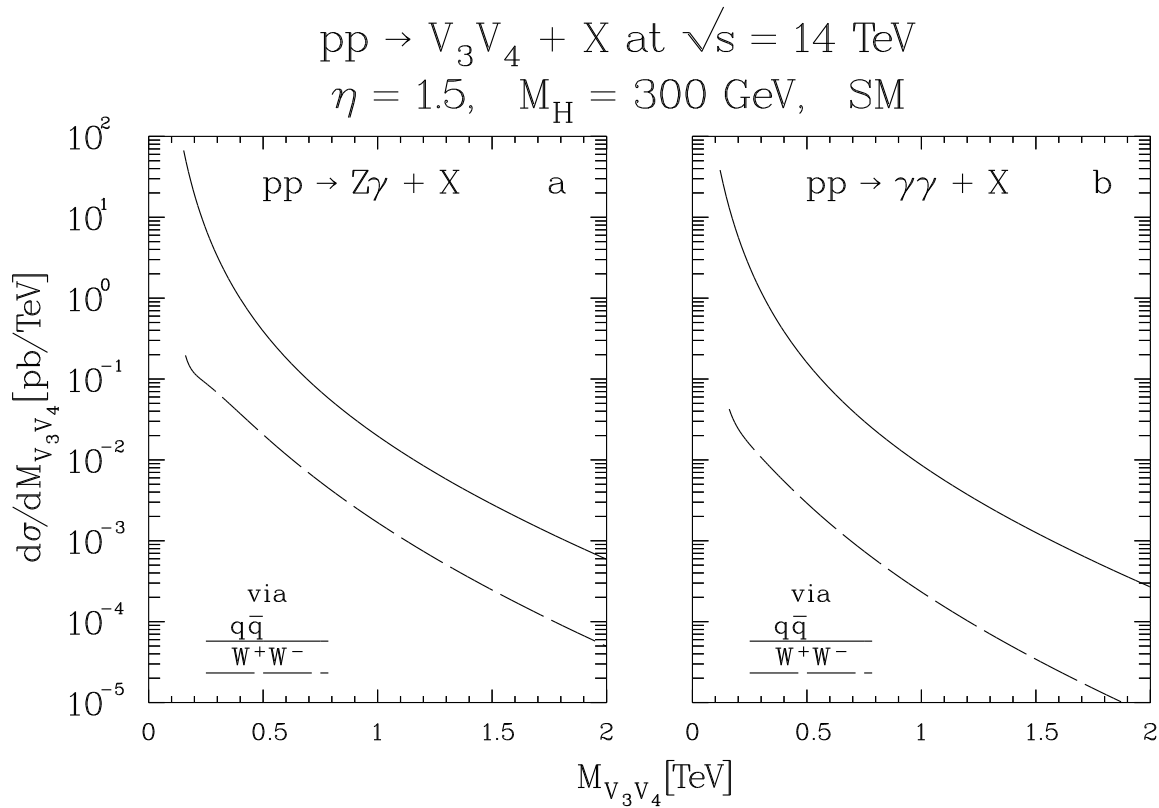,width=16cm,height=11cm}
\end{center}
\caption{The same as Fig. \ref{fig2} but for $Z\gamma$ and $\gamma\gamma$
production.}
\label{fig3}
\end{figure}

Figs.\ \ref{fig1},\ref{fig2}
and \ref{fig3} show the invariant mass distributions
for all vector boson pair production processes in the standard model.
We separately show the contributions from the processes $V_1V_2\to V_3V_4$ 
and those from $q\bar{q}'\to V_3V_4$.
The mass of the Higgs boson was chosen to be $M_H=300$ GeV.
There is little effect (less than 15\% of change in the contribution
from vector boson fusion) on the results for $W^\pm Z$ and $W^\pm\gamma$
production if the mass of the Higgs boson
is varied in between $M_H=0.1$ TeV and $M_H=0.8$ TeV.
The other electroweak
parameters were chosen as $\alpha=1/128, M_Z=91.19$ GeV and $M_W=80.33$ GeV.
We use the Higgs
boson width $\Gamma_H$ for the dominant decay modes into $W^+W^-$
and $ZZ$, $\Gamma_H=\Gamma_{H\to W^+W^-}+\Gamma_{H\to ZZ}$, where
\begin{equation}
\Gamma_{H\to VV}=\frac{\alpha M_H^3}{n!64s_W^2M_W^2}\sqrt{1-x_V}
(4-4x_V+3x_V^2)\label{gamma_h}.
\end{equation}
In (\ref{gamma_h}), $x_V\equiv 4M_V^2/M_H^2$ and $s_W^2\equiv
\sin^2(\theta_W)$, where $\theta_W$ is the weak mixing angle.
$n$ is the number of identical particles in the state $VV$.
For the parton distribution functions we use the set MRS(R2) \cite{mrsr}
which
includes the latest HERA data and uses $\alpha_s(M_Z^2)=0.120$ as
input parameter, a value favored
by the LEP~1 data. 
A contribution from top quarks is neglected.
For the scales $Q_i^2$ appearing as arguments 
of the parton distribution functions
we use the quark-quark sub-energy, 
$Q_i^2=s_{qq}$\footnote{If Approximation 2 of \cite{kuss} is
used for the vector-boson distribution functions, $Q^2=xs_{pp}$ has to be 
chosen instead, where $x$ is the first argument of $f_q^p(x,Q^2)$.}.
For the elements of the CKM matrix we take
$|V_{ud}|^2=|V_{cs}|^2=0.95$, $|V_{us}|^2=|V_{cd}|^2=0.05$ and consequently
assume
no mixing of the
third flavor with the other two flavors.
If no CKM mixing is included at all none of the differential
cross sections changes by 
more than 1\%.

Figs.\ \ref{fig1} to \ref{fig3} clearly
show that the contribution from vector boson scattering is
always an order of magnitude smaller than the contribution
from $q\bar{q}'$-annihilation (also if the sum over all $V_1V_2$ is taken).
The contribution may therefore indeed be neglected.

\begin{figure}
\begin{center}
\epsfig{file=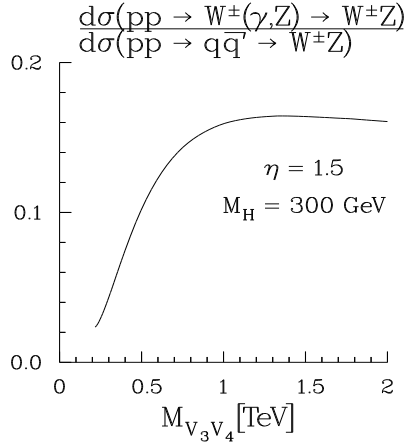,width=6cm,height=6cm}
\end{center}
\caption{The ratio of the cross sections for $pp\to W^{\pm}(Z,\gamma)\to
W^{\pm}Z$ and $pp\to q\bar{q}'\to W^{\pm}Z$ as a function of
the invariant mass $M_{V_3V_4}$ for $\protect\sqrt{s_{pp}}=14$ TeV.}
\label{rpl}
\end{figure}

Fig. \ref{rpl} shows the ratio of the cross sections for
$pp\to W^{\pm}(\gamma,Z)\to W^{\pm}Z$ ($W^\pm Z$ production
$\equiv$ the sum of $W^+Z$ and $W^-Z$ production, $W^\pm(\gamma,Z)$ 
intermediate states $\equiv$ the sum of $W^\pm\gamma$ and $W^\pm Z$
intermediate states)
and for $pp\to q\bar q'\to W^\pm Z$ 
as a function of $M_{V_3V_4}$.
The ratio of the integrated cross sections\footnote{For this numerical
evaluation we used $M_H=0.1$ TeV in order to be able to compare with
results in the literature.
We integrated the cross sections between 0.5 TeV $< M_{WZ} <$ 2 TeV.}
is 12\% (15\%) for a cut of $\eta=1.5$ ($\eta=2.5$).

\clearpage

\clearpage

We note that a different value of this ratio is obtained if the EVBA in leading
logarithmic approximation (LLA) is used instead.
In \cite{dobado1,dobado} the cross sections for $pp\to W^{\pm}(\gamma,
Z)\to W^{\pm}Z$ 
and for $pp\to q\bar{q}'\to W^{\pm}Z$ 
were calculated and the LLA EVBA was used.
Calculating the ratio of these cross sections, we obtain
57\% for $Y=1.5$ and 64\% for $Y=2.5$ for the case of a light Higgs
boson (59\% ($Y=1.5$) and 65\% ($Y=2.5$) for $M_H=1$ TeV).
Likewise, if we repeat the calculation of
\cite{renard,amps} (we used
$\eta=1.5$, $M_H=0.1$ TeV and integrated the cross sections in the region
0.5 TeV $<M_{WZ}<$ 2 TeV),
we obtain a value of 52\% for the ratio. For more details we refer to
\cite{diss}.

These values of the ratio are thus much higher than the values
obtained with the improved EVBA. The latter values are however
in agreement with values following from
\cite{goldplate}, in which a 
complete (lowest order) calculation of the processes
$q_1 q_2\to q'_1 q'_2 W^+ Z$ was carried out instead of an EVBA.
Computing the ratio of the cross sections for $pp\to q_1q_2\to q_1'q_2'
W^+Z$ and $pp\to q\bar{q}'\to W^+Z$ given in \cite{goldplate}
one obtains 17\% (21\%) for $M_H=0.1$ TeV ($M_H=1$ TeV).
In summary, the improved EVBA calculation and the 
complete calculation
both yield a value for the ratio which is between 10\% and
about 20\%,
while calculations using the LLA EVBA yield 
a value which is larger by more than a factor of 3.

We remark that for $M_H=300$ GeV even the Higgs boson
peak (which is present only in $W^+W^-$ and $ZZ$ production)
stays below the rate of $q\bar{q}'$-annihilation.
We finally note that the like-charge pair production processes 
$pp\to W^{\pm}W^{\pm}+X$ cannot
proceed via $q\bar{q}'$ annihilation and might thus allow
to directly observe vector boson scattering.

\subsection{Parametrization of Anomalous Couplings}
The model we use for the anomalous couplings was described in
\cite{gids} (GIDS model).
In this model, the most general $SU(2)_L\times U(1)_Y$-symmetric interaction
terms of dimension six are added to the Lagrangian of the standard model.
We restrict ourselves to $C$- and $P$-conserving interactions which contain
no higher derivatives and
explicitly contain vector boson self-interactions. There are three 
of those interaction terms which are
described by the parameters $\ew,\ewp$
and $\ebp$\footnote{The parameters are called $\epsilon_{W},\epsilon_{W\Phi}$
and $\epsilon_{B\Phi}$ in \cite{gids}.}. 
They are related to the usual parameters \cite{bkrs}
$x_{\gamma},y_{\gamma}$ and $\delta_Z,x_Z,y_Z$,
which parametrize the $C$- and 
$P$-conserving interactions of the $\gamma W^+W^-$ and the $Z W^+W^-$
vertex, respectively, by 
\[\delta_Z=\frac{c_W}{s_W}\Delta g_1^Z=\frac{\ewp}{s_Wc_W},\;\,
x_Z=\frac{c_W}{s_W}(\Delta\kappa_Z-\Delta g_1^Z)=
-\frac{s_W}{c_W}(\ewp+\ebp)=-\frac{s_W}{c_W}x_\gamma,\]
\begin{equation}
y_Z=\frac{c_W}{s_W}\lambda_Z=\frac{c_W}{s_W}\ew,\quad
x_{\gamma}=\Delta\kappa_\gamma=\ewp+\ebp,\quad
y_{\gamma}=\lambda_\gamma=\ew.
\label{couprel}\end{equation} 
In (\ref{couprel}) we also included the relations to the parameters
$\Delta g_1^Z,\Delta\kappa_Z,\lambda_Z$ and
$\Delta\kappa_{\gamma},\lambda_{\gamma}$ of \cite{zep}.

The reduction from the five parameter case of
$\delta_Z,x_Z,y_Z,x_\gamma$ and $y_\gamma$ to the three parameter case
is manifest through the relations $x_Z=-(s_W/c_W)x_\gamma$ and
$y_Z=(c_W/s_W)y_\gamma$ which are implicit in (\ref{couprel}).
The three parameter model defined in (\ref{couprel}) has already been
obtained \cite{gids} in \cite{kmss2,kmss} from the assumption of a
custodial $SU(2)$ symmetry.  
The relation between $x_{\gamma}$ and $x_Z$ in (\ref{couprel}),
$x_Z=-(s_W/c_W)x_{\gamma}$, is a consequence of the exclusion of
intrinsic $SU(2)$ violation, i.e., of $SU(2)$ custodial symmetry.
The relation between $y_{\gamma}$ and $y_Z$
in (\ref{couprel})
follows from the requirement of $SU(2)_L\times U(1)_Y$ symmetry
in the quadrupole interactions.
In addition to trilinear interactions the three-parameter dimension-six
$SU(2)_L\times U(1)_Y$ gauge invariant model
describes interactions among
four and more vector bosons. Also these interactions are already
contained in an identical form \cite{gids}
in the model described in \cite{kmss2,kmss}.
The only difference
\cite{gids2} of the three-parameter model 
\cite{kmss2,kmss} and the $SU(2)_L\times U(1)_Y$ 
invariant one lies
in non-standard interactions of the Higgs boson.

We note that there are no non-standard interactions among three {\em
neutral}
gauge bosons which would obey $C$- and $P$-symmetry, contain no higher
derivatives and are compatible with electromagnetic gauge and Lorentz
invariance \cite{peccei}. 

The Lagrangian of the GIDS model is an effective, unrenormalizable one
and can in general be written as \cite{georgi}
\begin{equation}
{\cal L}_{\mathrm{eff}}={\cal L}_0+\sum_j\frac{\tilde{g_j}}{\Lambda}
{\cal L}_j^{(5)}
+\sum_j\frac{\tilde{g_j}}{\Lambda^2}{\cal L}_j^{(6)}+\ldots\;.
\label{lexpand}
\end{equation}
In (\ref{lexpand}), ${\cal L}_0$ is the Lagrangian of the standard
model,
the ${\cal L}_j^{(d)}$ are interaction terms of dimension $d$, 
the $\tilde{g_j}$ are coupling constants 
and $\Lambda$ is the energy scale of new physics. We assumed the same
(gauge) symmetries for the ${\cal L}_j^{(d)}$ as for ${\cal L}_0$.
This implies that the ${\cal L}_j^{(5)}$ term (and all terms with an odd
$d$) in (\ref{lexpand}) are absent.
If we further assume that the
$\tilde{g}_j$ are of the same order of magnitude as the standard model
couplings $g$,$g'$ and $e$ and compare the Lagrangian
(\ref{lexpand}) with the one defining the $\alpha$-parameters \cite{gids},
we read off the order  
of magnitude for the $\alpha$-parameters,
\begin{equation}
\ew,\ewp,\ebp=O\left(M_W^2/\Lambda^2\right).
\end{equation}
Assuming $\Lambda\gsim 2$ TeV (and consequently restricting
ourselves to scattering energies up to $M_{V_3V_4}<2$ TeV),
the order of magnitude for the $\alpha$-parameters is
\begin{equation}
\ew,\ewp,\ebp\lsim {O}\left(10^{-3}\right).
\label{expected}\end{equation}

The restrictions derived from partial wave unitarity
applied to vector boson scattering amplitudes are \cite{uni}:
\begin{equation}
\left|\frac{s\ew}{M_W^2}\right|\lsim \sqrt{\frac{12s_W^2}{\alpha}}
\simeq 19,\quad \left|\frac{s\ewp}{M_W^2}\right|\lsim 15.5,
\quad\left|\frac{s\ebp}{M_W^2}\right|\lsim 49,
\label{unitarity}\end{equation}
where we have introduced $s\equiv M_{V_3V_4}^2$.
For $\sqrt{s}\le 2$ TeV the unitarity bounds (\ref{unitarity}) are
\begin{equation}
|\ew|\le 0.031,\quad\quad |\ewp|\le 0.025,\quad\quad
|\ebp|\le 0.079.
\label{alphalimits}\end{equation}
These limits are larger than the values in Eq.\ (\ref{expected})
for the $\alpha's$ which we expect from the effective Lagrangian ansatz.
Therefore, if the couplings are not larger than expected from the
effective Lagrangian
ansatz, unitarity is not violated for energies $\sqrt{s}\le 2$ TeV.
In \cite{zep,berger,baur,bho_ww}
a form factor assumption is made in order to avoid
violation of unitarity.
In our fits we follow the simple prescription to vary the coupling
parameters within their unitarity limits only. 
In fact it will turn out that within the 95\% CL limits the
unitarity limits are never reached.
Thus, in order to derive sensible experimental bounds on the anomalous
couplings, one does not have to use form factors for which
additional (unknown) parameters must be introduced.

If one nevertheless introduces
a form factor, the couplings $\alpha_i$ which are to be
inserted in the expressions for the cross sections are energy dependent.
They are related to bare (energy independent) coupling constants,
$\alpha_i^{0}$, by
\begin{equation}
\alpha_i = \frac{\alpha_i^{0}}{\left(1+\frac{\textstyle s}{
\textstyle\Lambda_{FF}^2}\right)^{\textstyle n}}.
\label{ff}\end{equation}
The bare coupling constants are those which appear in the Lagrangian.
A usual choice for the exponent $n$ in
(\ref{ff}) is $n=2$. Similar to $\Lambda$ in (\ref{lexpand}) $\Lambda_{FF}$ is
an energy scale for new physics.
The unitarity limits for the parameters $\alpha_i^{0}$ are obtained by
inserting (\ref{ff}) into (\ref{unitarity}). We use $n=2$ and minimize
the maximum value for $|\alpha_i^0|$ with respect to $s$. The minimum
occurs at $s=\Lambda_{FF}^2$ and the unitarity limits are given by
\[|\ew^0|\lsim 76(M_W^2/\Lambda_{FF}^2)\simeq 0.123,\quad\quad
|\ewp^0|\lsim 62(M_W^2/\Lambda_{FF}^2)\simeq 0.100,\]
\begin{equation}
|\ebp^0|\lsim 196(M_W^2/\Lambda_{FF}^2)\simeq 0.316.
\label{12}\end{equation}
The numerical values in (\ref{12}) are for $\Lambda_{FF}=2$ TeV.
At multi-TeV colliders the cross section for fixed 
$\alpha_i^{0}\neq 0$ is very different from the cross section for
fixed $\alpha_i$ and the obtainable bounds on the $\alpha_i$ are
very much tighter than those for the $\alpha_i^{0}$.
The distinction between the two models
does however not very much affect the analysis of 
present Tevatron data 
since there the form factor is close to the value 1
as $\sqrt{s}$ can hardly be greater than 0.5 TeV. 

\subsection{Results with Anomalous Couplings}

\begin{figure}
\begin{center}
\epsfig{file=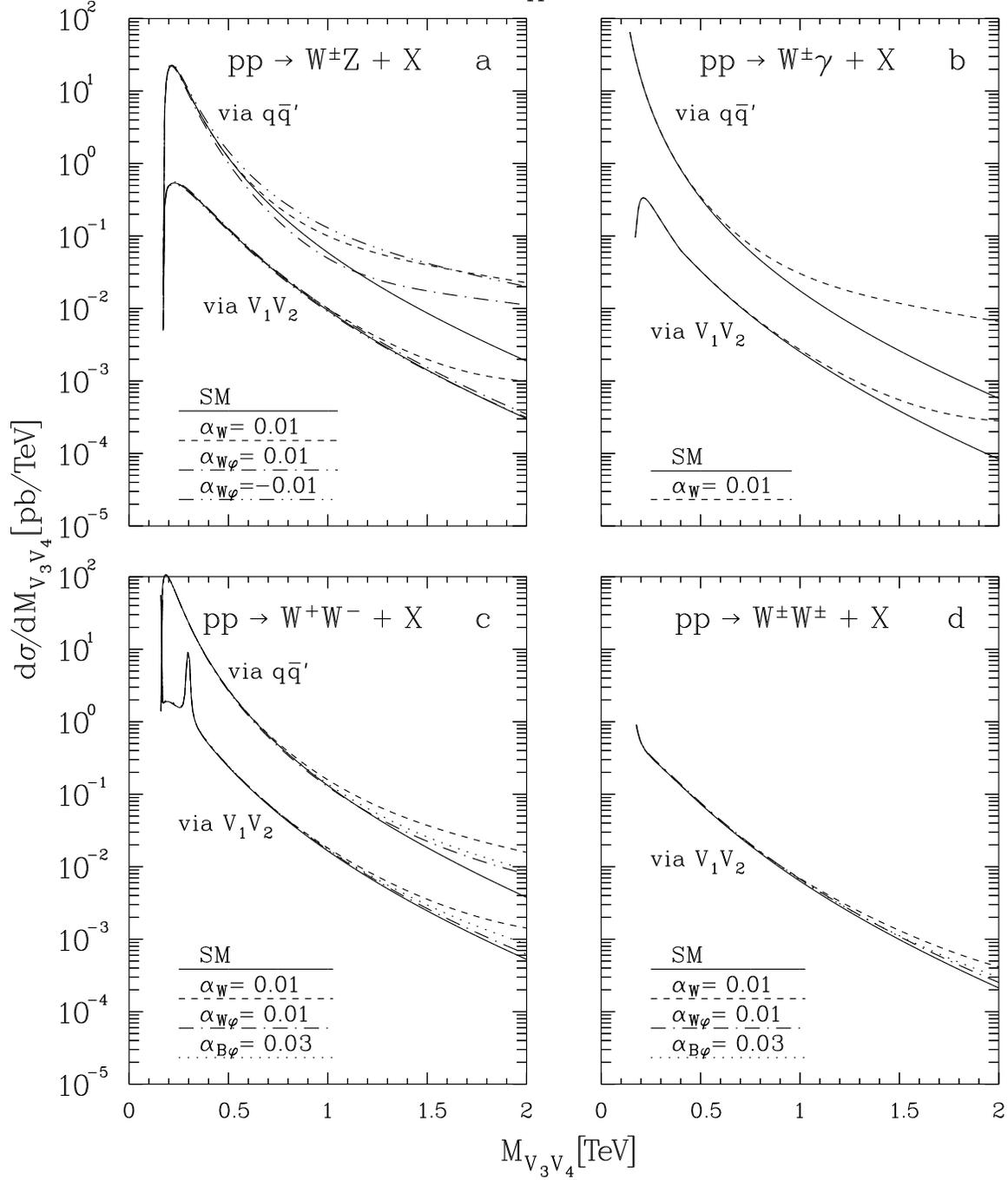,width=16cm,height=20cm}
\end{center}
\caption{
The cross-sections for $W^{\pm}Z (\equiv W^+Z+W^-Z), 
W^{\pm}\gamma, W^+W^-$ and
$W^{\pm}W^{\pm} (\equiv W^+W^++W^-W^-)$ 
production as functions of the invariant mass $M_{V_3V_4}$ of the produced
vector boson pair for $pp$-collisions
at $\protect\sqrt{s}=14$ TeV. Various values of
the anomalous couplings have been chosen. Separately
shown are the contribution from $q\bar q'$-annihilation and the (summed)
contribution from vector boson fusion processes.
A pseudorapidity cut of $\eta=1.5$ has been applied.}
\label{fig4}
\end{figure}

Fig.\ \ref{fig4} shows the comparison of $q\bar{q}'$-annihilation and
vector boson fusion
in the presence of anomalous couplings. 
We show the results for the relevant processes of $W^\pm Z$ and
$W^\pm\gamma$ production and for $W^+W^-$ production.
In addition, we present a plot for $W^{\pm}W^{\pm}$ production.
We sum over the charge conjugated final
states
i.e. discuss the cross sections for $W^{\pm}V\equiv W^+V+W^-V$ and
$W^{\pm}W^{\pm}\equiv W^+W^++W^-W^-$ production. We have also summed
over all $V_1V_2$ pairs.
We only vary one coupling at a time. Only those 
couplings which lead to enhanced terms at high energies
(i.e. of ${O}(\alpha_i s)$ or
${O}(\alpha_i^2 s^2)$) in the $q\bar q'$ cross section are varied.
Varying the other couplings leaves the $q\bar q'$ cross sections
virtually unchanged.
For $W^\pm W^\pm$ production we vary all couplings.
We choose a single non-zero magnitude for each of the couplings
which is already quite large for
the effective Lagrangian expectation,
(\ref{expected}), but which is still below the unitarity limit
(\ref{unitarity}). For $\ew$ and $\ewp$ we take
$|\alpha_i|=0.01$. For $\ebp$ we take $|\ebp|=0.03$.
For the relevant processes of $W^{\pm}V$ production, we choose a negative and a
positive value for the coupling if there is an enhanced term linear in the
coupling. 

The main conclusion from Fig.\ \ref{fig4} is that vector boson scattering
is only marginally important even if the anomalous couplings are different from zero.
When constraining anomalous couplings using
these processes, vector boson scattering might therefore well be omitted.
The non-enhanced terms
($\ebp$ in $W^{\pm}Z$-production, $\ewp$ and
$\ebp$ in $W^{\pm}\gamma$-production)
are unlikely to lead to any observable effect at the LHC.
Fig. \ref{fig4} (d) shows that the effect of anomalous couplings for
like-charge $W^{\pm}$-pair production is not very large.

\section{Parameter Fits for Anomalous Couplings\label{fits}}
In this section we present parameter fits to fictitious standard model data
and derive limits for the anomalous couplings.
Refering to the conclusion of Section \ref{compsec},
we will take into account only the contribution from $q\bar{q}'$ annihilation.
First we consider $W^\pm\gamma$ and $W^\pm Z$ production separately.
These are the experimentally relevant production processes \cite{kuijf}. 
The detection of a $W^+W^-$ pair is experimentally
plagued by a large background of $t\bar{t}$ production with the
subsequent decay of a top quark into a $W^{\pm}$ boson and a $b$ quark
\cite{kuijf}.
We use the general parametrization of the triple gauge boson vertices
\cite{peccei,bkrs,laymou}
in terms of seven free parameters, thus allowing for $C$- and
$P$-violation.
Then we present a fit to combined $W^\pm\gamma$ and $W^\pm Z$ ``data'' for the
three parameter gauge invariant model.
We take into account the full correlations among the
parameters. Before we proceed we present the unitarity limits for the
set of couplings which we are using \cite{laymou,nuss}. As far as we
know, these limits have never been given before.

\subsection{Unitarity Limits for \boldmath $\delta_V,x_V,y_V,z_V,z'_{1V},
z'_{2V}$ and $z'_{3V}$\unboldmath}
Theoretical bounds on anomalous couplings can be obtained by
applying partial wave unitarity to the amplitudes for $q\bar{q}'\to
V_3V_4$.
Inequalities derived from the requirement of partial wave unitarity have
been given in \cite{unibaur}. The inequalities have been written
in terms of ``reduced amplitudes'' 
for $q\bar{q}'\to W^{\pm}Z$ and $q\bar{q}'\to W^{\pm}\gamma$ 
scattering.
The reduced amplitudes have been given in terms of the
parameters $g_1^V,\kappa_V,\lambda_V,g_4^V,g_5^V,\tilde{\kappa}_V$ and
$\tilde{\lambda}_V$, where $V=\gamma$ or $Z$. 
By comparison of the Lagrangians of \cite{laymou} and \cite{unibaur} we
find the following equivalence between this set of parameters and the
one we are using:
\begin{eqnarray}
g_1^V-1&=&\frac{\delta_V}{g_V^{SM}},\quad\quad
\lambda_V=\frac{y_V}{g_V^{SM}},\quad\quad
\kappa_V-1=\frac{\delta_V+x_V}{g_V^{SM}}\cr
g_4^V&=&\frac{z'_{1V}}{g_V^{SM}},\quad\quad
g_5^V=-\frac{z_V}{g_V^{SM}}\frac{M_V^2}{M_W^2}
+i\frac{z'_{3V}}{g_V^{SM}}\frac{(P^2-M_{W^*}^2)}{M_W^2},\quad\quad
\tilde{\lambda}_V=-2\frac{z'_{3V}}{g_V^{SM}}\cr
\tilde{\kappa}_V&=& -\frac{z'_{2V}}{g_V^{SM}}
-\frac{z'_{3V}}{g_V^{SM}}\frac{(P^2+M_{W^*}^2)}{M_W^2}
-i\frac{z_V}{g_V^{SM}}\frac{(P^2-M_{W^*}^2)}{M_W^2}.
\label{trans}\end{eqnarray}
In (\ref{trans}), $P^2$, $M_{W^*}^2$ and $M_V^2$ are the squared
invariant masses of the $W^+$, the $W^-$ and the $V$,
respectively, entering the trilinear vertex
and $g_{\gamma}^{SM}=1$, $g_Z^{SM}=(c_W/s_W)$.
Because the two parameter sets are only equivalent up to possible 
form factors\footnote{Form factors can be introduced by adding terms with
two or a larger even number of
derivatives on the fields to a Lagrangian with constant
couplings. These terms are equal to a power of a squared
invariant mass (or even a product of powers of several squared invariant
masses) times the interaction term of the original Lagrangian.
In order to compare the interaction terms of the Lagrangians of \cite{laymou} 
and \cite{unibaur} the terms of one Lagrangian have to be
re-grouped (by using partial integrations and tensor identities).
Two derivatives on a field appear in some of the re-grouped terms.
This introduces the $P^2$, $M_{W^*}^2$ and $M_V^2$ dependences in
(\ref{trans}).},
the equations (\ref{trans}) contain the kinematic variables $P^2,
M_{W^*}^2$ and $M_W^2$.
We checked that 
with the replacements (\ref{trans}) the expressions for the amplitudes
for $W^\pm\gamma,W^\pm Z$ and $W^+W^-$ production in terms of the two
sets translate in the correct way.

The following table summarizes the symmetry properties of the parameters 
under $C$ and $P$ transformations:

\begin{center}
\begin{tabular}{|c|c|c|c|}\hline
$\delta_V,x_V,y_V$ & $z_V$ & $z'_{1V}$ & $z'_{2V},z'_{3V}$\\ \hline
$C$,$P$ & $\not\! C$,$\not\! P$ ($CP$) & $\not\! C$, $P$ & $C$,$\not\! P$\\ 
\hline\end{tabular}
\end{center}

\noindent
If electromagnetic gauge invariance is demanded the following parameters 
vanish,
\begin{equation}
U(1)_{\mathrm{e.m.}}\quad\rightarrow\quad \delta_\gamma = 0,\quad
z'_{1\gamma}=0.
\label{u1em}
\end{equation}

Assuming that only one anomalous coupling at a time is different from zero
we extract the unitarity bounds shown in Tables \ref{uni2} and
\ref{uni1} from the bounds on 
the reduced amplitudes
in \cite{unibaur}. For $W^{\pm}Z$ production we neglected terms
of ${O}(M_W^2/s)$.
For the form factor case we used (\ref{ff}) with $n=2$ and minimized the
unitarity bounds with respect to $s$. However, for $z_\gamma^0$, $z_Z^0$
and $(z'_{3Z})^0$ the value of $s$ at the minimum is greater than
$\Lambda_{FF}^2$. For these cases we quote the unitarity
limit for $s=\Lambda_{FF}^2$.
The bounds shown in Tables \ref{uni2} and \ref{uni1} are weaker than
those derived from vector boson scattering
because in the latter processes
the amplitude is in general quadratic in the couplings while for 
$q\bar q'\to V_3V_4$ it is at most linear.

\begin{table}\begin{center}
\renewcommand{\arraystretch}{2}
\begin{tabular}{|c|l|c||c|l|c|}\hline
Para- & Unitarity & $\sqrt{s}=$ & Para- & Unitarity & $\Lambda_{FF}=$\\
meter & limit & 2 TeV & meter & limit & 2 TeV \\ \hline
$|\delta_{\gamma}|$ & $K/(\sqrt{2}\nu)\simeq 6.0/\sqrt{s}, s\gg M_W^2$ &
3.0 & $|\delta_\gamma^0|$ & $96/(3\sqrt{3}\Lambda_{FF})$ & 9.2 \\
$|x_{\gamma}|$ & $\sqrt{2}K/\nu\simeq 12.0/\sqrt{s}$ & 6.0 &
$|x_\gamma^0|$ & $192/(3\sqrt{3}\Lambda_{FF})$ & 18.5 \\
$|y_{\gamma}|$ & $\sqrt{2}KM_W/(\sqrt{s}\nu)\simeq 0.96/s$ & 0.24 &
$|y_\gamma^0|$ & $3.84/\Lambda_{FF}^2$ & 0.96 \\
$|z_{\gamma}|$ & $\sqrt{2}K/[\nu((s/M_W^2)-1)]\simeq 0.077/s^{3/2}$ & 
 0.0096 & $|z_\gamma^0|$ & $0.308/\Lambda_{FF}^3$ & 0.039 \\
$|z'_{1\gamma}|$ & $\sqrt{2}K/\nu\simeq 12.0/\sqrt{s}$ & 6.0 &
$|(z'_{1\gamma})^0|$ & $192/(3\sqrt{3}\Lambda_{FF})$ & 18.5 \\
$|z'_{2\gamma}|$ & $\sqrt{2}K/\nu\simeq 12.0/\sqrt{s}$ & 6.0 &
$|(z'_{2\gamma})^0|$ & $192/(3\sqrt{3}\Lambda_{FF})$ & 18.5 \\
\hline
\end{tabular}\end{center}
\caption{Unitarity limits for $\gamma W^+W^-$ couplings without and with
a form factor derived
from partial wave
unitarity for $q\bar{q}'\to W^{\pm}\gamma$. $K=\sqrt{3}s_W/(\alpha
\sqrt{1-M_W^2/s})$, $\nu=\sqrt{s/M_W^2+1}$.}
\label{uni2}
\end{table}

\begin{table}\begin{center}
\renewcommand{\arraystretch}{2}
\begin{tabular}{|c|l|c||c|l|c|}
\hline
Para- & Unitarity & $\sqrt{s}=$ & Para- & Unitarity & $\Lambda_{FF}=$ \\ 
meter & limit & 2 TeV & meter & limit & 2 TeV \\ \hline
$|\delta_Z|$ & $(2K/s_W)(M_W^2/s)\simeq 1.54/s$ & 0.39 &
$|\delta_Z^0|$ & $6.16/\Lambda_{FF}^2$ & 1.54 \\
$|x_Z|$ & $\sqrt{2}K(c_W/s_W)(M_W/\sqrt{s})\simeq 12.0/\sqrt{s}$ & 6.0 &
$|x_Z^0|$ & $192/(3\sqrt{3}\Lambda_{FF})$ & 18.5 \\
$|y_Z|$ & $\sqrt{2}K(c_W/s_W)(M_W^2/s)\simeq 0.96/s$ & 0.24 &
$|y_Z^0|$ & $3.84/\Lambda_{FF}^2$ & 0.96 \\
$|z_Z|$ & $\sqrt{2}K(c_W/s_W)(M_W^3/s^{3/2})\simeq
 0.077/s^{3/2}$ & 0.0096 & $|z_Z^0|$ & $0.308/\Lambda_{FF}^3$ &
0.039 \\
$|z'_{1Z}|$ & $(2K/s_W)(M_W^2/s)\simeq 1.54/s$ & 0.39 &
$|(z'_{1Z})^0|$ & $6.16/\Lambda_{FF}^2$ & 1.54 \\
$|z'_{2Z}|$ & $\sqrt{2}K(c_W/s_W)(M_W/\sqrt{s})\simeq 12.0/\sqrt{s}$ & 6.0 &
$|(z'_{2Z})^0|$ & $192/(3\sqrt{3}\Lambda_{FF})$ & 18.5 \\
$|z'_{3Z}|$ & $K/(\sqrt{2}s_W)(M_W^3/s^{3/2})\simeq
 0.044/s^{3/2}$ & 0.0055 & $|(z'_{3Z})^0|$ &
$0.176/\Lambda_{FF}^3$ & 0.022 \\ \hline
\end{tabular}\end{center}
\caption{Unitarity limits for $ZW^+W^-$ couplings without and with a
form factor derived from partial wave
unitarity for $q\bar{q}'\to W^{\pm}Z$. $K=\sqrt{3}s_W^2/(\alpha c_W)$.}
\label{uni1}
\end{table}

\subsection{Present Direct Limits}
At present, direct limits on the couplings have been obtained
by the CDF and D0 collaborations at Tevatron \cite{tevatron,wood}
and by the LEP~2 collaborations \cite{lep97,mele}.
Table \ref{tevalimits} summarizes the most stringent bounds which were
attained at the Tevatron.
\begin{table}
\begin{center}
\begin{tabular}{|c|c|c|c|c|}
\hline
diboson pair&$W\gamma$&$W^+W^-/WZ$&$W^+W^-$&$W^+W^-/WZ$\\ \hline
assumptions&none&\multicolumn{2}{c|}{$\delta_Z=0,x_Z=\frac{c_W}{s_W}
x_{\gamma},y_Z=\frac{c_W}{s_W}y_{\gamma}$}&$y_\gamma=y_Z=0,x_\gamma=0$\\
\hline
&$-1.4<x_\gamma<1.4$&$-0.4<x_\gamma<0.6$&
$-0.9<x_\gamma<1.0$&$-2.5<\delta_Z<2.7$\\
\raisebox{1.5ex}[-1.5ex]{results}&$-0.5<y_\gamma<0.5$&
$-0.4<y_\gamma<0.4$&$-0.7<y_\gamma<0.7$&$-4<x_Z<4$\\ \hline
$\Lambda_{FF}$/TeV&1.5&2&2&1\\ \hline
\end{tabular}
\end{center}
\caption{Results from \cite{tevatron,wood} for the 95\% CL limits
on anomalous couplings obtained from two-parameter fits to data 
of diboson production in $p\bar{p}$ collisions at $\sqrt{s}=1.8$ TeV. 
The bounds take
into account possible correlations between the two fitted parameters.
The $C$- or $P$-violating couplings were assumed to be zero.
The value of $\Lambda_{FF}$ which was used in the fits is indicated in
the bottom row.}
\label{tevalimits}
\end{table}

The LEP~2 collaborations recently gave \cite{mele}
a preliminary limit for $\ewp$,
\[-0.3<\ewp<0.4,\quad\quad 95\%\;\mathrm{CL},\]
where $\ebp=\ew=0$ was assumed. 
Adopting a two-parameter model \cite{ks} which is equivalent to
$\ewp,\ebp\ne 0$ and $\ew=0$,
the following preliminary limits were obtained at LEP~2 \cite{mele},
\[|\delta_Z|<1.9,\quad\quad -2.5<x_{\gamma}<3.8,\quad\quad 95\%\;
\mathrm{CL}.\]
These limits take into account the correlations between the 
two parameters. No form factor was used.

The final sensitivity of LEP~2 has been estimated in \cite{bkrs,LEP2}.
For the three parameter gauge invariant model the following result
was obtained for a run at
$\sqrt{s}=190$ GeV with an integrated luminosity of ${\cal L}=500~
\mathrm{pb}^{-1}$ \cite{bkrs}: 
\[-0.20<\ew<0.24,\quad\quad -0.19<\ewp<0.13,\quad\quad -0.35<\ebp<1.05.\]
These bounds are at 95\% CL and take into account all correlations.

\subsection{Fitting Procedure}

We performed fits to the $M_{VV}$ and $p_T$ distributions
of the cross sections, where $p_T=q\sin\theta$ is the transverse 
momentum of a produced vector boson. The $p_T$
distribution was calculated according to
\begin{eqnarray}
&&\frac{d\sigma}{dp_T}
(h_1h_2\to q\bar q'\to V_3V_4,s_{hh})|_{\mathrm{cut}}\cr
&=&\frac{1}{p_T}\int\limits_{x_{\mathrm{min}}}^{x_{\mathrm max}}dx
\frac{(p_T/q)^2}{z_T} 
\int\limits_{\max[(1/2)\ln x,-y_0(z_T)]}^{\min[-(1/2)\ln x,y_0(z_T)]}
\!\!\!\!\!dy
\sum_{q\bar q'}
\left[f_q^{h_1}(\sqrt{x}e^y,Q_1^2)f_{\bar q'}^{h_2}(\sqrt xe^{-y},Q_2^2)
\;\;+\;\;h_1\leftrightarrow h_2\;\right]\cr
&&\times\left[\frac{d\sigma}{d\cos\theta}(q\bar q'\to V_3V_4,z_T)
+\frac{d\sigma}{d\cos\theta}(q\bar q'\to V_3V_4,-z_T)\right].
\label{dsdpt}
\end{eqnarray}
In (\ref{dsdpt}), $z_T\equiv \sqrt{1-(p_T/q)^2}$ is the magnitude of
$\cos\theta$ for the given $p_T$
and $x_{\mathrm{min}}$ and $x_{\mathrm max}$ are determined by
\begin{eqnarray}
x_{\mathrm{min}}&=&\max\left[(M_3+M_4)^2/s_{hh},x(p_T^2)\right],\cr
x_{\mathrm max}&=&\min\left[1,x(p_T^2/\sin^2\vartheta_{\mathrm min}),
(2\;\mbox{TeV})^2/s_{hh}
\right].\label{xminmax}
\end{eqnarray}
In (\ref{xminmax}) we included the upper bound of 2 TeV for $\sqrt{s}$.
$y_0(z_T)$ in (\ref{dsdpt}) is determined by the pseudorapidity cut.
In the high-energy limit ($q^2\gg
M_{3,4}^2$) it is given by 
\begin{equation}
y_0(z_T)\simeq \eta - \tanh^{-1}(z_T).
\end{equation} 
The function $x(q^2)$ in (\ref{xminmax}) is given by
\begin{equation}
x(q^2)=\frac{2q^2+M_3^2+M_4^2+2\sqrt{q^4+q^2(M_3^2+M_4^2)+M_3^2M_4^2}}
{s_{hh}}
\end{equation}
and $\vartheta_{\mathrm min}$ in (\ref{xminmax}) is determined by the
relation
$\tanh(\eta)=\cos\vartheta_{\mathrm min}$. 
$q$ is the variable defined in (\ref{zlimits}).

\begin{figure}
\begin{center}
\epsfig{file=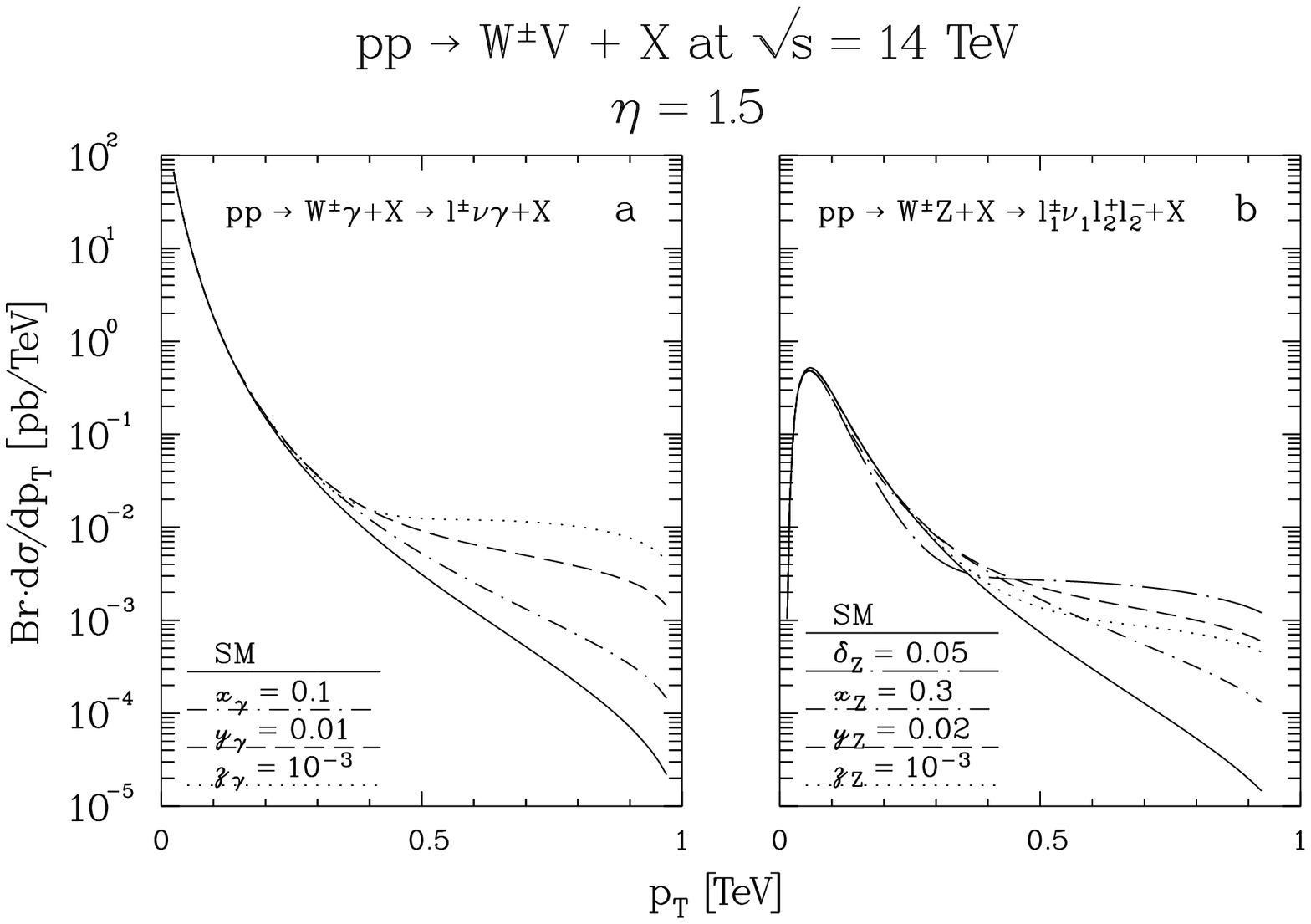,width=14.5cm,height=10cm}
\end{center}
\caption{The cross sections for $pp\to W^{\pm}\gamma+X$ (a) and 
$pp\to W^{\pm}Z+X$ (b) as a function of the transverse momentum
$p_T$ of a produced vector boson
in the standard model and for various values of the anomalous
couplings. The cross sections have been multiplied by the branching
ratios for the decays of massive vector bosons into two generations of
leptons.}
\label{ptpl}
\end{figure}

We assumed that the $W^{\pm},Z$ particles are identified by
their decays into two generations of leptons each. We used the
following branching ratios,
\begin{center}
\begin{tabular}{|r@{$\to$}l|r||r@{$\to$}l|r|}\hline
$W^{\pm}$&$l^{\pm}\nu$&10.8\%&$Z$&$l^+l^-$&
3.37\%\\ \hline
\end{tabular}
\end{center} 
We used no other cut than $\eta=1.5$ on the produced bosons. 
Figs.\ \ref{ptpl} (a) and (b) show the cross sections 
for $pp\to W^{\pm}\gamma+X$ and $pp\to W^{\pm}Z+X$, respectively,
multiplied by the branching ratios
as a function of $p_T$ in the standard model and for various values
of the anomalous couplings. No form factor was used.

To estimate the number of events at the LHC
we assume an integrated luminosity of ${\cal L}=10^5~
\mathrm{pb}^{-1}$.
We arrange fictitious standard model data into bins.
For the $M_{VV}$ distribution for $pp\to W^{\pm}Z+X$
we find that
there is less than $1$ event for
$M_{VV}>2$ TeV and $\simeq 21$ events in the interval 1 TeV $<M_{VV}<$
2 TeV. We choose this interval to be the first bin. The other bins
and numbers of SM events for $W^\pm Z$ and $W^\pm\gamma$ production
are shown in Table \ref{tabcoef}. 
We also show the numbers of events for $ZZ,Z\gamma$ and $\gamma\gamma$ 
production\footnote{We neglected potential contributions from gluon fusion
\cite{gluon}.}. 
The accuracy of the numbers due to numerical integration is 1\%.
We proceed to arrange the data for the $p_T$ distributions into bins.
Since $p_T\simeq M_{VV}/2$ for scattering at right angles and large
invariant masses we choose the limits for the $p_T$ bins equal to half
the limits of the $M_{VV}$ bins. 
In addition we define an eighth bin. The results are also shown in
Table \ref{tabcoef}.

\begin{table}
\hrule
\begin{center}
\begin{tabular}{|c|c|r|r|r|r|r|r|r|}\hline
\multicolumn{2}{|r|}{Bin Nr.}&1&2&3&4&5&6&7\\ \hline
\multicolumn{2}{|r|}{$M_{VV}$ [TeV]}&1-2&0.8-1&0.7-0.8&0.6-0.7&
0.5-0.6&0.4-0.5&0.3-0.4\\ \hline
&$W^{\pm}\gamma$&95&126&134&247&499&$1154$&$3320$\\
\cline{2-9}
$N_{SM}$&$W^{\pm}Z$&21&29&31&59&121&$285$&$823$\\ \cline{2-9}
$=$&$ZZ$&3.3&4.7&5.2&9.8&20.4&49&147\\ \cline{2-9}
${\cal L}\cdot Br\cdot\sigma$&$Z\gamma$&33&45&48&89&182&426&$1247$
\\ \cline{2-9}
&$\gamma\gamma$&214&287&304&556&$1110$&$2518$&
$6965$\\
\hline
\end{tabular}
\end{center}
\begin{center}
\begin{tabular}{|c|c|r|r|r|r|r|r|r|r|}\hline
\multicolumn{2}{|r|}{Bin Nr.}&1&2&3&4&5&6&7&8\\ \hline
\multicolumn{2}{|r|}{$p_T$ [TeV]}&0.5-1&0.4-0.5&0.35-0.4&0.3-0.35&
0.25-0.3&0.2-0.25&0.15-0.2&0.1-0.15\\ \hline
&$W^{\pm}\gamma$&32&51&55&103&208&473&$1275$&
$4603$\\ \cline{2-10}
&$W^{\pm}Z$&7.6&12&13&24&48&108&275&821\\ \cline{2-10}
$N_{SM}$&$ZZ$&1.6&2.5&2.7&4.9&9.8&21.5&53&151\\ \cline{2-10}
&$Z\gamma$&17&25&27&50&101&225&595&$2048$\\ \cline{2-10}
&$\gamma\gamma$&116&172&187&345&$698$&$1600$
&$4460$&$17600$\\ \hline
\end{tabular}
\end{center}
\caption{The numbers of standard events in 7 bins over the invariant
mass $M_{VV}$ and in 8 bins over the transverse momentum
$p_T$ for the processes
$pp\to V_3V_4+X$ at $\sqrt{s}=14$ TeV with a cut
$|\eta|<1.5$ on the pseudorapidity of the produced vector bosons
and the requirement $\sqrt{s}<2$ TeV for the $p_T$ distributions.
An integrated
luminosity of ${\cal L}=10^5~\mathrm{pb}^{-1}$ has been assumed.
For massive vector bosons a decay 
into two generations of leptons was assumed.
All results were obtained in the Born approximation.~~~~~~~~~~
~~~~~~~~~~~~~~~~~~~~~~~~~~~~~~~~~~~~~~~~}
\label{tabcoef}
\hrule
\end{table}

To calculate the non-standard effects we wrote
the number of events in each bin as a power
series in the anomalous couplings,
\begin{equation}
N(\alpha's)=N_{SM}+\sum_i\alpha_i N_i+\sum_{i,j}\alpha_i\alpha_j N_{ij},
\label{coefdef}
\end{equation}
where the $\alpha_i$ are the anomalous couplings.
Table \ref{coeftab} shows the coefficients for the $C$- and $P$-conserving
couplings and for $z_\gamma$ 
in a bin of the $p_T$ distribution comprising 0.4 TeV $<p_T<$ 1 TeV.
\begin{table}
\hrule
\begin{center}
\begin{tabular}{|c|r|r|r|r|r|r|r|r|r|r|}
\hline
&SM&$\hat x_\gamma$&$\hat y_\gamma$&$\hat z_\gamma$&$\hat x^2_\gamma$&
$\hat y_\gamma^2$&$\hat z_\gamma^2$&$\hat x_\gamma\hat y_\gamma$&
$\hat x_\gamma\hat z_\gamma$&$\hat y_\gamma\hat z_\gamma$\\ \cline{2-11}
\raisebox{1.5ex}[-1.5ex]{$W^\pm\gamma$}&82&$-2.5$&0&$-6.1$&63.5&233&497&
31.8&0.1&0.01\\ \hline\hline
&SM&$\hat\delta_Z$&$\hat x_Z$&$\hat y_Z$&$\hat\delta_Z^2$&$\hat x_Z^2$&
$\hat y_Z^2$&$\hat\delta_Z\hat x_Z$&$\hat\delta_Z\hat y_Z$&
$\hat x_Z\hat y_Z$\\ \cline{2-11}
\raisebox{1.5ex}[-1.5ex]{$W^\pm
Z$}&20&$-12.7$&$-3.0$&$-0.68$&6.3&4.1&15.3&2.3&0.57&2.0\\ \hline
\end{tabular}
\end{center}
\caption{The coefficients $N_i$ and $N_{ij}$, defined
in Eq. (\ref{coefdef}), for the numbers of produced $W^\pm\gamma$ and $W^\pm Z$
pairs in a bin with 0.4 TeV $<p_T<$ 1 TeV
for the rescaled coupling parameters
$\hat x_\gamma\equiv x_\gamma\cdot 10$,
$\hat y_\gamma\equiv y_\gamma\cdot 100$,
$\hat z_\gamma\equiv z_\gamma\cdot 10^3$ and 
$\hat\delta_Z\equiv\delta_Z\cdot 100$,
$\hat x_Z\equiv x_Z\cdot 10$, 
$\hat y_Z\equiv y_Z\cdot 100$. The numbers of standard events are also
shown. Sample usage: For $y_\gamma=2\cdot 10^{-3}\Leftrightarrow
\hat y_\gamma = 0.2$ and all other anomalous couplings equal to zero there are
$82+233\cdot (0.2)^2\simeq 91$ $W^\pm\gamma$ events in the bin,
corresponding to one standard deviation from the standard model.
\label{coeftab}}
\hrule
\end{table}
 
In each bin we calculate 
\begin{equation}
\Delta\chi^2=-2\ln\left(\frac{\cal L}{{\cal L}_0}\right),
\end{equation}
where ${\cal L}$ is the likelihood function for the data in this bin,
assuming that a theory with particular (non-zero) 
values for the anomalous couplings is the correct theory,
and ${\cal L}_0$ is the same function assuming that the standard model is
correct. 
$\Delta\chi^2$ is a measure of the probability that this particular
model can still describe the (standard) data.
If the number of events (in the bin) is greater than 50,
we calculate ${\cal L}$ according to a Poisson distribution of the total
number of events,
\begin{equation}
{\cal L}=p_N=\frac{<N>^N}{N!}e^{-<N>}.\label{poisson}
\end{equation}
In (\ref{poisson}), $<N>\equiv N(\alpha's)$ is the number of events
predicted by the particular non-standard theory and $N\equiv N_{SM}$
is the number of standard events (=the number of ``measured''
events). 
If the number of events is smaller than 50 we generate the $N$
events in the bin, i.e. we calculate the phase space
points $\Omega_i,i=1\ldots N$, at which the standard events would be located in
the bin. $\Omega$ represents $M_{VV}$ or $p_T$ for the two
distributions, respectively.
We then use the method of
extended maximum likelihood (EML) to calculate ${\cal L}$.
The likelihood function of the EML is given by
\begin{equation}
{\cal L}_{EML}=p_N\prod_i^N p(\Omega_i,\vec{\alpha}),
\end{equation}
with $p_N$ from (\ref{poisson}) and $p(\Omega_i,\vec{\alpha})$ is
the probability of finding the $i$th event at the phase space point
$\Omega_i$, assuming that the theory with the parameters $\vec{\alpha}$
is correct. $p(\Omega_i,\vec{\alpha})$
is given in terms of the differential cross section by
\begin{equation}
p(\Omega_i,\vec{\alpha})=\frac{1}{\sigma}\frac{d\sigma}{d\Omega}(\Omega_i),
\quad\mathrm{with}\quad
\sigma=\int_{\mathrm{bin}}\frac{d\sigma}{d\Omega}d\Omega,
\end{equation}
where $\sigma$ and $d\sigma/d\Omega$ are evaluated in the non-standard
theory.

\subsection{Results}

\begin{figure}
\begin{center}
\epsfig{file=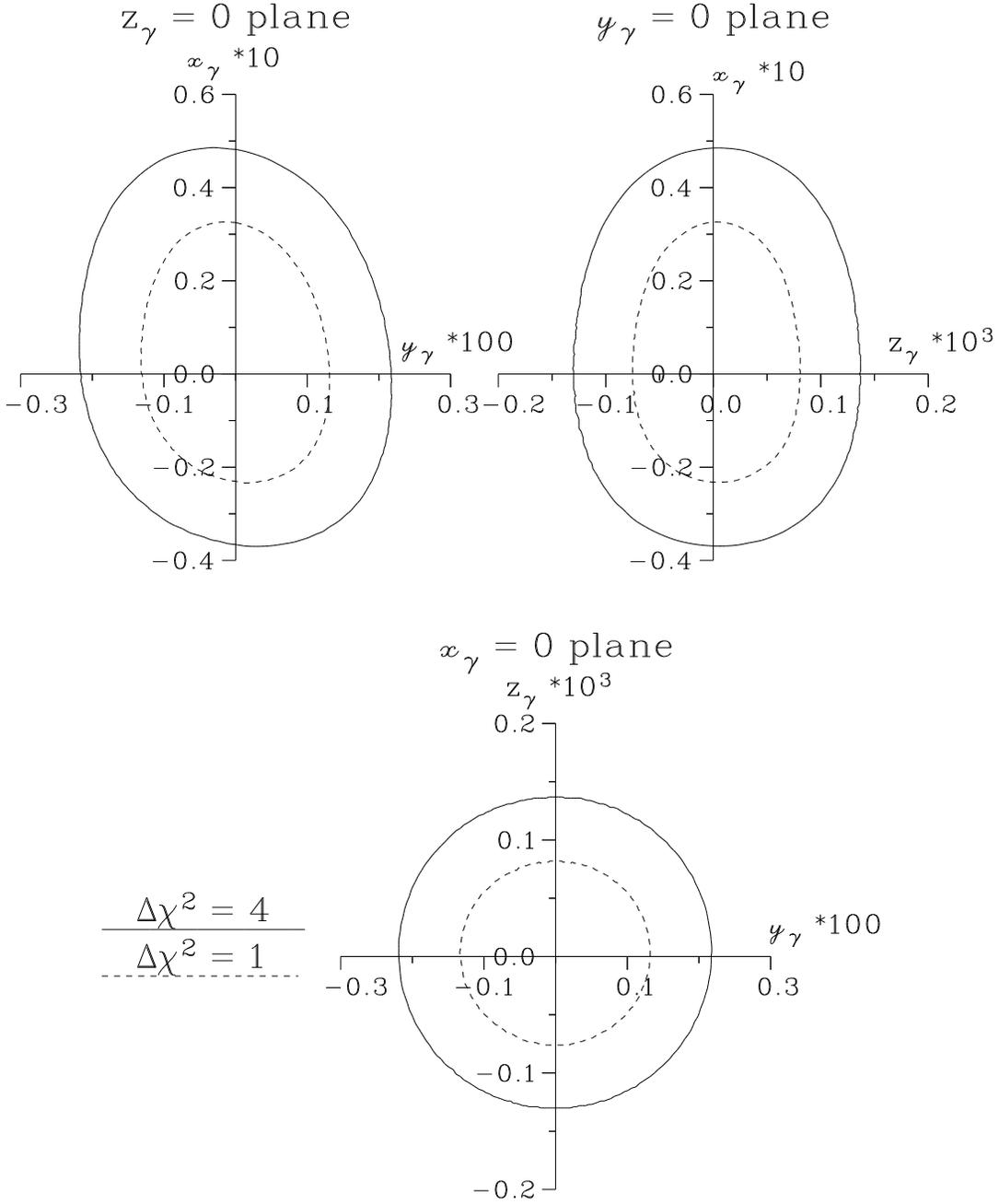,width=14.5cm,height=20cm}
\end{center}
\caption{The projections of the $\Delta\chi^2=4$ and $\Delta\chi^2=1$
confidence regions on the $x_{\gamma},y_{\gamma}$ and $z_{\gamma}$ 
parameter planes from a three parameter
fit of the $p_T$
distribution of $pp\to W^{\pm}\gamma+X$ at $\sqrt{s}=14$ TeV with a cut
of $\eta=1.5$. An integrated luminosity of ${\cal L}=10^5\;
\mathrm{pb}^{-1}$ and a leptonic decay of the $W^{\pm}$ boson
into two generations of fermions was assumed. All other parameters were
assumed to be equal to zero.}
\label{wg1fit}
\end{figure}

Fig.\ \ref{wg1fit} shows the projections of the $\Delta\chi^2=1$ and
$\Delta\chi^2=4$ confidence regions on the 
$x_{\gamma}=0,y_{\gamma}=0$ and $z_{\gamma}=0$ parameter planes
as the result of a three parameter fit to the $p_T$ distribution of $pp\to
W^{\pm}\gamma+X$.
The parameters $z'_{1\gamma},z'_{2\gamma}$ and $z'_{3\gamma}$ were set
equal to zero.
As the parameters are
uncorrelated, the projections are equal to the sections of the
confidence regions with the planes.
If the $M_{VV}$ distribution is used instead, the regions
expand by a factor of 1.1 to 1.15 in each dimension. 
If only the total number of events
in each bin is subjected to the fit (instead of using the EML method),
the regions expand by a factor of about 1.2 in each dimension.
If a four parameter fit to $x_\gamma,y_\gamma,z_\gamma$ and
$z'_{2\gamma}$ is performed instead, the projections and sections stay
the same as in Fig.\ \ref{wg1fit}. The parameter $z'_{3\gamma}$ does not
contribute to $W^\pm\gamma$ production. We assumed
$\delta_\gamma=z'_{1\gamma}=0$ because of electromagnetic gauge
invariance.

\begin{figure}
\begin{center}
\epsfig{file=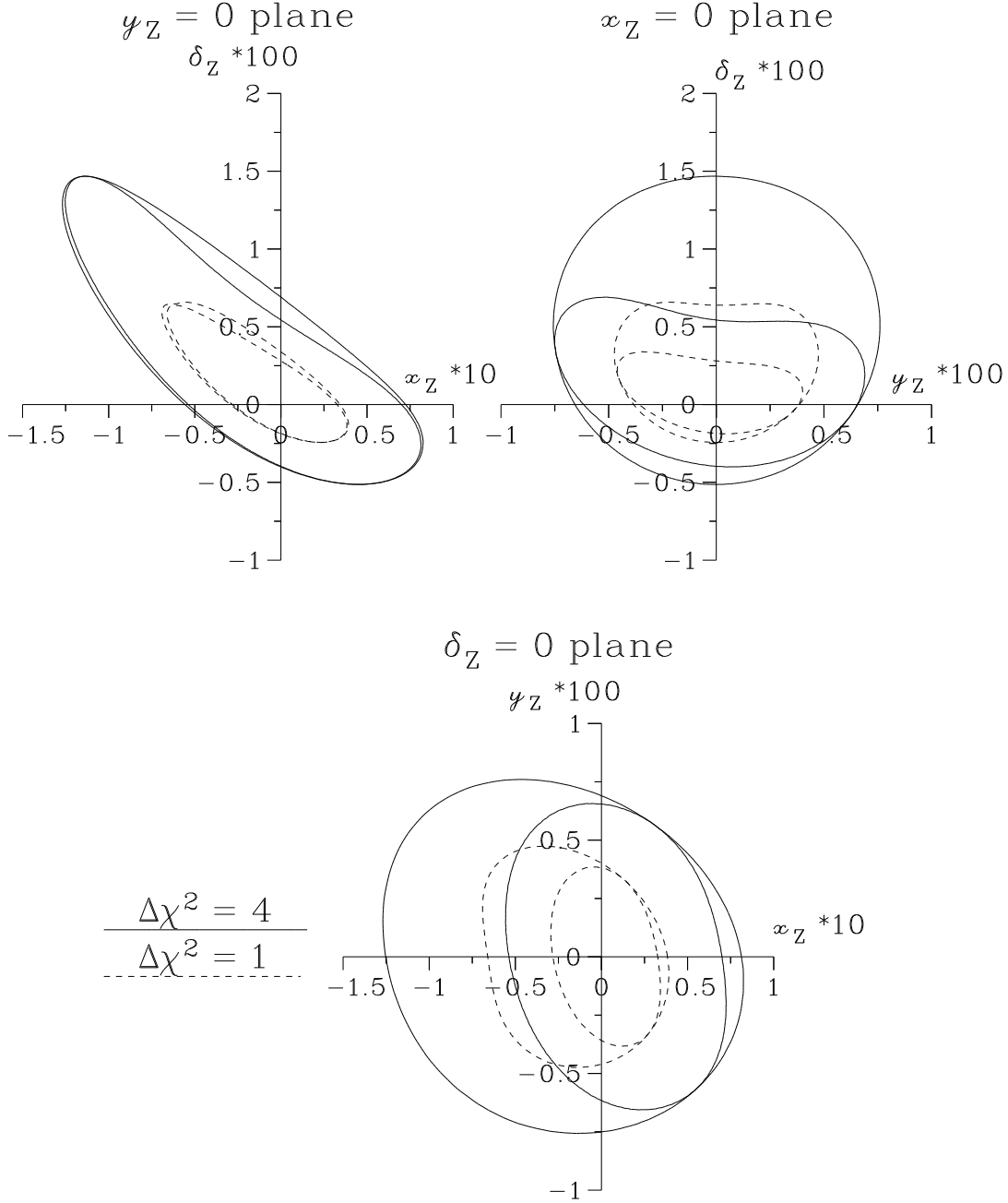,width=14.5cm,height=20cm}
\end{center}
\caption{The projections of the $\Delta\chi^2=4$ and $\Delta\chi^2=1$
confidence regions on the $\delta_Z,x_Z$ and $y_Z$ parameter planes and the
sections of the regions with these planes from a three parameter
fit of the $p_T$
distribution of $pp\to W^{\pm}Z+X$ at $\sqrt{s}=14$ TeV with a cut
of $\eta=1.5$. An integrated luminosity of ${\cal L}=10^5\;
\mathrm{pb}^{-1}$ and a leptonic decay of the $W^{\pm},Z$ bosons
into two generations of fermions was assumed. All other parameters were
assumed to be equal to zero. The sections are drawn 
in the same way as the projections. They can be distinguished from the
projections as they always lie inside them.}
\label{wz1fit}
\end{figure}

Fig.\ \ref{wz1fit} shows the projections of the confidence regions on the
parameter planes $\delta_Z=0$,
$x_Z=0$ and $y_Z=0$ and the sections of the regions with the planes
as the result of a three parameter fit to the $p_T$
distribution of $pp\to W^{\pm}Z+X$.
The parameters $z_Z,z'_{1Z},z'_{2Z}$ and $z'_{3Z}$ were set equal to
zero.
The figure displays the correlations among the parameters. As a result of
the correlations the sections are smaller than the projections.
A four parameter fit which includes $z_Z$ yields identical results.
If a seven parameter fit (including also $z'_{1Z},z'_{2Z}$ and $z'_{3Z}$)
is performed, the projected confidence region expands by no more than
4\% in any direction except for the positive $\delta_Z$ direction for
$\chi^2=1$ where it expands by $\simeq 30\%$.

\begin{figure}
\begin{center}
\epsfig{file=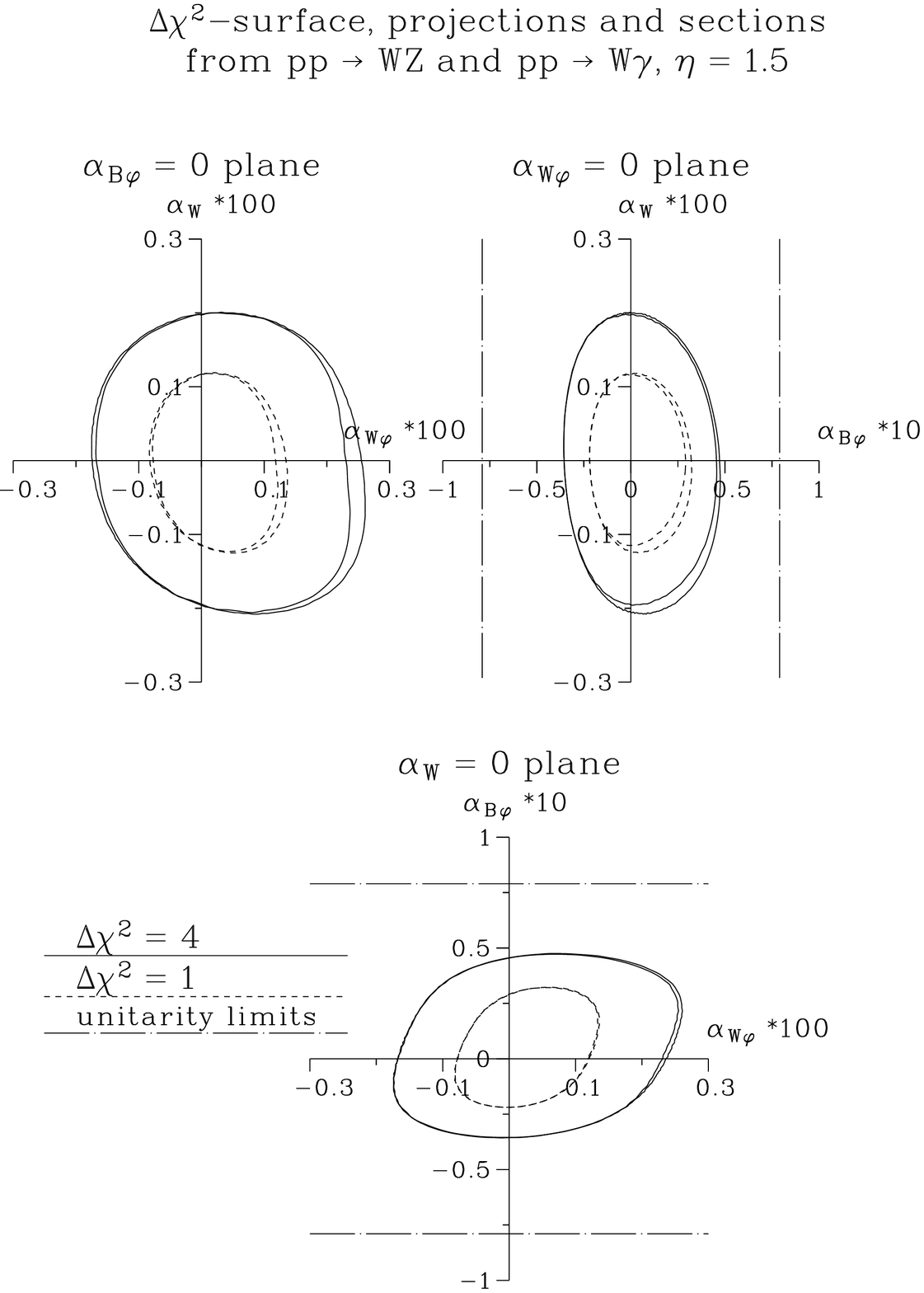,width=14.5cm,height=20cm}
\end{center}
\caption{The projections of the $\Delta\chi^2=4$ and $\Delta\chi^2=1$
confidence regions on the $\ew,\ewp$ and $\ebp$
parameter planes and the
sections of the regions with these planes from a simultaneous
fit of the $p_T$
distributions of $pp\to W^{\pm}Z+X$ and $pp\to W^{\pm}\gamma+X$
to the three parameter gauge invariant model
for $\sqrt{s}=14$ TeV and a cut
of $\eta=1.5$. 
The unitarity limits are also shown.
An integrated luminosity of ${\cal L}=10^5\;
\mathrm{pb}^{-1}$ and a leptonic decay of the $W^{\pm},Z$ bosons
into two generations of fermions was assumed. All other parameters 
were assumed to be equal to zero. 
The sections are drawn in the same way as the projections. 
They can be distinguished from the projections as they
always lie inside them.}
\label{epsfit}
\end{figure}

Figure \ref{epsfit} shows the projections and sections on
the $\ew=0,\ewp=0$ and $\ebp=0$ planes from a simultaneous fit
of the $p_T$ distributions of $pp\to W^{\pm}Z+X$ and
$pp\to W^{\pm}\gamma+X$ to the three parameter gauge invariant model.
The unitarity limits for the parameters
are also shown. 
The confidence regions lie inside the unitarity limits.

The use of different parton distribution
functions leads to small theoretical
uncertainties ($<1$\%) in the confidence regions. These uncertainties
could be reduced by subjecting ratios of cross sections,
e.g. $\sigma(W^\pm V)/\sigma(\gamma\gamma)$, 
to the fit. 
Due to the additional statistical error induced by the reference
cross section (i.e. $\sigma(\gamma\gamma)$ in our example) 
the confidence levels derived from the ratios
are, however, several tens of percent
wider than those derived from the absolute values of the cross sections. 
We do not use ratios for our fits.

We repeat our analyses using a form factor.
We project the confidence regions on the parameter axes. 
This results in 95\% (for $\chi^2=4$) and 68\% ($\chi^2=1$) confidence
limits for the parameters.
Table \ref{results} summarizes our results. If we repeat the fits for
$\eta=3$ the bounds are only slightly affected: the differences 
between the maximal and minimal values of the couplings
change by at most 20\% compared to Table \ref{results}.
In general the differences decrease.

\begin{table}
\begin{center}
\begin{tabular}{|c||r|r||r|r|||c||r|r||r|r|}
\hline
\multicolumn{5}{|c|||}{no form factor}&
\multicolumn{5}{c|}{with form factor}\\
&\multicolumn{2}{c||}{68\% CL}&\multicolumn{2}{c|||}{95\% CL}&
&\multicolumn{2}{c||}{68\% CL}&\multicolumn{2}{c|}{95\% CL}\\
\raisebox{1.5ex}[-1.5ex]{parameter}&min&max&min&max&
\raisebox{1.5ex}[-1.5ex]{parameter}&min&max&min&max\\
\hline
$\delta_Z\cdot 100$&$-0.24$&0.87&$-0.51$&1.48&
$\delta_Z^0\cdot 100$&$-0.49$&3.40&$-0.99$&4.41\\
$x_Z\cdot 10$&$-0.68$&0.38&$-1.26$&0.82&
$x_Z^0\cdot 10$&$-1.77$&0.64&$-2.76$&1.29\\
$y_Z\cdot 100$&$-0.46$&0.46&$-0.75$&0.76&
$y_Z^0\cdot 100$&$-1.44$&1.43&$-1.93$&1.94\\
$z_Z\cdot 10^3$&$-0.26$&0.26&$-0.45$&0.45&
$z_Z^0\cdot 10^3$&$-0.85$&0.84&$-1.30$&1.29\\
$z'_{1Z}\cdot 100$&$-0.74$&0.74&$-1.21$&1.21&
$(z'_{1Z})^0\cdot 100$&$-2.8$&2.8&$-3.1$&3.1\\
$z'_{2Z}\cdot 10$&$-1.35$&1.35&$-1.79$&1.79&
$(z'_{2Z})^0\cdot 10$&$-3.0$&3.0&$-3.4$&3.4\\
$z'_{3Z}\cdot 10^4$&$-1.47$&1.47&$-2.55$&2.55&
$(z'_{3Z})^0\cdot 10^4$&$-4.9$&4.9&$-7.4$&7.4\\
\hline
$x_{\gamma}\cdot 10$&$-0.23$&0.33&$-0.37$&0.49&
$x_{\gamma}^0\cdot 10$&$-0.31$&0.56&$-0.49$&0.77\\
$y_{\gamma}\cdot 100$&$-0.131$&0.131&$-0.22$&0.22&
$y_{\gamma}^0\cdot 100$&$-0.35$&0.34&$-0.52$&0.50\\
$z_{\gamma}\cdot 10^3$&$-0.075$&0.081&$-0.129$&0.135&
$z_{\gamma}^0\cdot 10^3$&$-0.23$&0.25&$-0.37$&0.39\\
$z'_{2\gamma}\cdot 10$&$-0.30$&0.30&$-0.44$&0.44&
$(z'_{2\gamma})^0\cdot 10$&$-0.45$&0.45&$-0.63$&0.63\\
\hline
$\ew\cdot 100$&$-0.125$&0.119&$-0.21$&0.20&
$\ew^0\cdot 100$&$-0.34$&0.31&$-0.51$&0.47\\
$\ewp\cdot 100$&$-0.082$&0.136&$-0.175$&0.26&
$\ewp^0\cdot 100$&$-0.145$&0.22&$-0.29$&0.42\\
$\ebp\cdot 10$&$-0.22$&0.32&$-0.36$&0.48&
$\ebp^0\cdot 10$&$-0.29$&0.53&$-0.47$&0.74\\
\hline
\end{tabular}
\end{center}
\caption{The projections of the $\Delta\chi^2=1$ (68\%
CL) and $\Delta\chi^2=4$ (95\% CL) confidence regions on the parameter
axes as the results of a seven parameter fit of the $p_T$ distribution
of $pp\to W^\pm Z+X$ to $\delta_Z,x_Z,y_Z,z_Z,
z'_{1Z},z'_{2Z}$ and $z'_{3Z}$,
a four parameter fit of the $p_T$ distribution of $pp\to W^\pm\gamma+X$
to $x_\gamma,y_\gamma,z_\gamma$ and $z'_{2\gamma}$
and a three parameter fit of the combined $p_T$
distributions of $pp\to W^\pm Z+X$ and $pp\to W^\pm\gamma +X$ to
$\ew,\ewp$ and $\ebp$. The form
factor results are for $\Lambda_{FF}=2$ TeV and $n=2$.}
\label{results}
\end{table}

The 95\% confidence limits which we obtain 
for the alternative set of parameters 
$\Delta g_1^Z,\Delta\kappa_Z$ and $\lambda_Z$ of \cite{zep} (instead of
$\delta_Z,x_Z$ and $y_Z$) are:
\begin{eqnarray}
-0.0028<\Delta g_1^Z<0.0080,&&-0.0052<\Delta g_1^{Z,0}<0.024,\cr
-0.062<\Delta\kappa_Z<0.044,&&-0.13<\Delta\kappa_Z^0<0.062,\cr
-0.0041<\lambda_Z<0.0041,&&-0.0103<\lambda_Z^0<0.0104.
\label{baurlimits}
\end{eqnarray}

We compare our results with previous investigations.
Sensitivity limits achievable at the LHC were
previously presented in \cite{falk,bho_wg,barger,baur,bho_ww}.
Fits to the $p_T$ distribution of 
fictitious data for $pp\to W^+Z+X$ at $\sqrt{s}=14$ TeV
were performed in \cite{baur}. 
The 95\% CL limits presented there,
using a form factor with $\Lambda_{FF}=3$ TeV and $n=2$
and based on the Born level prediction were 
\[
-0.0048 <\Delta g_1^{Z,0}< 0.0164,\quad
-0.120 <\Delta\kappa_Z^0< 0.092,\quad
-0.0082 <\lambda_Z^0< 0.0084.\]
If we repeat our three-parameter fit with $\Lambda_{FF}=3$ TeV
we obtain a similar result, namely\footnote{
Different cuts were used in \cite{baur}. In particular, a pseudorapidity
cut of $\eta=3$ was applied on the decay products of the vector bosons.
If we repeat our analysis for $\eta=3$ and include, as in \cite{baur},
only $W^+Z$ production our limits change by less than 3\%.}
\[ -0.0039 < \Delta g_1^{Z,0}< 0.0140,\quad
-0.090 <\Delta\kappa_Z^0< 0.053,\quad
-0.0067 <\lambda_Z^0< 0.0068.\]

An SSC analysis using a form factor for $pp\to W^+\gamma+X$ can be found in
\cite{bho_wg}. If we repeat our analysis with the parameters used in
\cite{bho_wg} ($\sqrt{s}=40$ TeV, $W^+$ decays to only one lepton family,
${\cal L}=10^4\;\mathrm{pb}^{-1}$ and using only the information
from the total number of events in each bin) and use a cut of $\eta=2.5$ we
obtain -0.17 $<\Delta\kappa_{\gamma}^0<$ 0.21 and
-0.021 $<\lambda_{\gamma}^0<$ 0.020 at 95\% CL\footnote{We 
required $p_T>200$ GeV.}.
These bounds are tighter by a factor of 1.5 to 2 than the ones obtained
in \cite{bho_wg}.

In \cite{barger}, an LHC bound on $x_{\gamma}$ was derived, assuming 
$y_{\gamma}=0$. This bound was derived from the $O(\alpha_s)$ prediction
for the cross section, but from Table IV of \cite{barger} 
we deduce that the 1$\sigma$ bound which would be
obtained from the Born approximation 
is $|x_{\gamma}|<0.06$\footnote{
We assumed a quadratic
dependence of the number of predicted non-standard events on
this coupling.} (we do not use a form factor for this comparison). 
This bound is wider by a factor of in between two and three than
our bound. This can be explained by the fact that in \cite{barger}
the assumed luminosity was only ${\cal L}=3\cdot 10^4\;
\mathrm{pb}^{-1}$, only $W^+\gamma$ production was considered,
the fitting procedure was simpler (only one bin was taken) and
different cuts were used. Including the ${O}(\alpha_s)$ corrections reduces the
sensitivity to $x_\gamma$ by about a factor of two \cite{bho_wg,barger}.

The limits which were derived in \cite{falk} are much larger than ours because
the discovery criterion employed there is much stronger than ours.
We note that 
the chiral Lagrangian parameters $x_9^L$ and $x_9^R$ used in \cite{falk}
are identical to $\ewp$ and $\ebp$, respectively\footnote{This is true 
as far as only the trilinear vector couplings are concerned.}.
The explicit connection is given by
\begin{equation}
\frac{\alpha}{8\pi s_W^2}x_9^L=-\ewp,\quad\quad
\frac{\alpha}{8\pi s_W^2}x_9^R=-\ebp.
\end{equation}

\section*{Conclusion}
We showed that the rate for vector boson fusion production 
of vector boson pairs at the LHC is at the order of
10\% to 20\% of quark antiquark annihilation production and 
might thus be neglected in an estimate of the pair production
cross sections.
This result was obtained by applying an improved formulation of the
effective vector boson approximation (EVBA). It agrees with the result
of a calculation in which the complete set of diagrams was evaluated
instead of performing an EVBA. Previous calculations in which the EVBA
in leading logarithmic approximation was used overestimate the
contribution from vector boson fusion by a factor of 3.

We derived confidence
intervals for the full set of anomalous $W^+W^-\gamma$ and $W^+W^-Z$
couplings, including
$C$- and $P$-violating couplings,
from fits to the standard $W^\pm\gamma$ and $W^\pm Z$ production rates
expected for the LHC.
In addition we derived confidence intervals for a three parameter 
$SU(2)_L\times U(1)_Y$ gauge
invariant dimension-six extension of the standard model.
We performed multi-parameter fits in which the full number of anomalous
couplings was varied.
Our limits thus take into account all possible correlations among the effects
of the various possible couplings. 
We derived limits with and without making a form factor assumption.
We compare the limits with the unitarity limits for
the production of vector boson pairs with invariant masses smaller
than $\sqrt{s}=2$ TeV.
It turns out that all 95\% confidence limits lie inside the unitarity limits 
whether a form factor is used or not.
It is therefore not necessary to use a form factor in order to
avoid violation of unitarity.
The limits which we obtain without using a form factor
are a factor of 10 (for $x_\gamma$ or
$\ebp$) and 100 (for ($\delta_Z,y_\gamma$) or ($\ewp,\ew$)) stronger than
the present experimental limits or limits which can be attained at LEP~2.

Adopting an effective Lagrangian approach, 
together with an assumption about the
energy scale at which new physics occurs, provides us with an order of 
magnitude estimate for the parameters $\ew$, $\ewp$ and $\ebp$.
We find that for $\ew$ and $\ewp$, the 
limits which can be obtained at the LHC
are of the same order of magnitude as this estimate.
It might thus be possible to observe non-zero
values of these coupling parameters, should they exist, at the LHC.

In appendices we give analytical expressions for cross sections for
vector boson pair production with anomalous couplings for
$q\bar q'$ annihilation and vector boson fusion. Our expressions
manifestly show the effects of the couplings at large scattering
energies.

\section*{Acknowledgments}
I.K. thanks D. Schildknecht for giving him the opportunity to do
this work and F.M. Renard for financial support during his visit in
Montpellier. 
We thank S. Dittmaier, J.-L. Kneur, J. Layssac,
G. Moultaka, F.M. Renard, D. Schildknecht
and H. Spiesberger for discussions and comments.

\begin{appendix}
\section{Cross-Sections for \boldmath $q\bar{q}'$-Annihilation
\unboldmath\label{appa}}
The standard model differential cross sections to ${O}(\alpha^2)$
for $q\bar{q}'\to W^{\pm}Z$,
$q\bar{q}'\to W^{\pm}\gamma$, $q\bar{q}\to W^+W^-$ and $q\bar{q}\to ZZ$
have been first given in \cite{brown}\footnote{Also the $x_{\gamma}$-terms 
for the $W^{\pm}\gamma$ production cross section have been given there.}.
In a form in which good high energy behavior is manifest and
including the $\ew$-interaction,
all cross sections for $q\bar{q}'\to V_3V_4$ can be found in \cite{renard}.
For arbitrary vector boson self-interactions
all cross sections and helicity amplitudes 
have been recently given in \cite{nuss}.

We give here
the formulas for the differential cross sections 
for the $q\bar q'$ processes which 
receive contributions from anomalous vector boson self-interactions,
$q\bar{q}'\to W^{\pm}Z, q\bar{q}'\to W^{\pm}\gamma$ and $q\bar{q}\to W^+W^-$,
in a form in which the high energy behavior is manifest.
As in \cite{renard}, we have explicitly carried out the 
high energy cancellations
among different diagrams, also (as far as possible) for the 
non-standard terms.   
We use the general $C$- and $P$-conserving vector boson self-interactions
compatible with Lorentz-invariance and electromagnetic gauge invariance
in terms of the parameters
$x_\gamma,y_\gamma,\delta_Z,x_Z$ and $y_Z$. 
In addition, we include the contributions from $z_\gamma$ and 
$z'_{2\gamma}$\footnote{The parameter $z'_{3\gamma}$ does not
contribute.} for $W^\pm\gamma$ production and the contributions from
$z_Z,z'_{1Z},z'_{2Z}$ and $z'_{3Z}$ for $W^\pm Z$ production.
The 
differential cross sections for $q\bar{q}\to ZZ,\gamma Z,\gamma\gamma$ 
can be found in \cite{renard}.

For $q\bar{q}'\to W^{\pm}V$, $V=\gamma,Z$, the cross sections contain an 
overall factor of $|V_{q\bar q'}|^2$, where $V_{q\bar q'}$ 
is the element of the CKM matrix for the
mixing of the quarks $q$ and $q'$. For $q\bar{q}\to W^+W^-$, the quarks have to
be of the same flavor.
We give the cross sections averaged over colors and spins of the initial quarks
and summed over the helicities of the final state vector bosons. 
The cross sections given here agree with the expressions given in
\cite{brown}, \cite{renard}\footnote{After correction of misprints.}
and \cite{nuss}. 

We denote the left- and right-handed couplings of the $Z$-boson to the 
quarks by
\begin{eqnarray}
L_u= 1-\frac{4}{3}s_W^2,&&\quad R_u=-\frac{4}{3}s_W^2\\
L_d=-1+\frac{2}{3}s_W^2,&&\quad R_d= \frac{2}{3}s_W^2.
\end{eqnarray}
We also use the symbols
\begin{equation}
\tau_3^u= 1,\quad \tau_3^d= -1;
\quad\quad Q_u=\frac{2}{3},\quad Q_d=-\frac{1}{3}.
\end{equation}
The Mandelstam variable $t$ will be defined below for each process and
$u$ is defined by $s+t+u=M_3^2+M_4^2$. 
The scattering angle $\theta$
is the angle between the three-momenta of the two
particles which define $t$. We further use the variables
\[ \beta\equiv\sqrt{1-\frac{2(M_3^2+M_4^2)}{s}+\frac{(M_3^2-M_4^2)^2}{s^2}}, \]
and $\eta$, where $\eta=\mp 1$ for $W^\pm V_4$ production. 
We treat the processes $q\bar{q}'\to W^{\pm}Z$ and 
$q\bar{q}'\to W^{\pm}\gamma$ together
because similar functions are involved.

The differential cross sections are given by:

\vspace{3mm}
\centerline{\underline{$q\bar{q}'\to W^{\pm}V_4$}}
\nopagebreak

~~~~~~~~~~1. \underline{$q\bar{q}'\to W^{\pm}\gamma$}
\nopagebreak

\begin{eqnarray}
\frac{d\sigma}{d\cos\theta}&=&\frac{\pi\alpha^2\beta}{24ss_W^4}|V_{q\bar
q'}|^2
\Bigg\{ 2s_W^2\left(\frac{1}{1+u/t}-\frac{1}{3}\right)^2
\cdot\left(\frac{s^2+M_W^4}{tu}-2\right)\cr
&&+x_{\gamma}\frac{s_W^2}{s-M_W^2}\left(\frac{4tu}{s-M_W^2}
+\frac{2}{3}(u+2t)\right)\cr
&&+\eta
s_W^2\frac{s}{M_W^2}z_{\gamma}\left(\frac{1}{3}-\cos\theta\right)
\left(1+\frac{M_W^2}{s}\right)\cr
&&+\frac{s_W^2}{4}z_{\gamma}^2\left(\frac{s(t^2+u^2)}{M_W^6}
+\frac{4tu}{M_W^4}\right)
+\eta\frac{s_W^2}{2}\beta\cos\theta
\frac{s^2}{M_W^4}z_{\gamma}(x_{\gamma}+y_{\gamma})\cr
&&+\frac{1}{2}\left(\frac{ss_W}{s-M_W^2}\right)^2\bigg[
y_{\gamma}^2A_{y^2}(0) +2x_{\gamma}y_{\gamma}A_{xy}(0)
+\left(x_{\gamma}^2+(z'_{2\gamma})^2\right)A_{x^2}(0)\bigg]\Bigg\},
\end{eqnarray}
where we have defined $t$ as
\begin{equation}
t=\left\{ \begin{array}{c} (p_u-p_{W^+})^2 \\
                           (p_{\bar u}-p_{W^-})^2
        \end{array}
\right.,\label{tdef}
\end{equation}
for the two charge conjugated processes, respectively. In, (\ref{tdef})
$p_i$ denotes the four-momentum of the particle $i$. 

\vspace{2mm}
~~~~~~~~~~2. \underline{$q\bar{q}'\to W^{\pm}Z$}
\nopagebreak

\begin{eqnarray}
\frac{d\sigma}{d\cos\theta}&=&\frac{\pi\alpha^2\beta}{24ss_W^4}|V_{q
\bar{q}'}|^2\Bigg\{
\frac{1}{(s-M_W^2)^2}\left[\frac{(9-8s_W^2)}{2}(tu-M_W^2M_Z^2) +(4s_W^2-3)s
(M_W^2+M_Z^2)\right]\cr
&&-\frac{2}{s-M_W^2}\left[tu-M_W^2M_Z^2-s(M_W^2+M_Z^2)\right]
\left(\frac{L_d}{t} -\frac{L_u}{u}\right)\cr
&&+\frac{(tu-M_W^2M_Z^2)}{2c_W^2}\left(\frac{L_d^2}{t^2}+\frac{L_u^2}{u^2}
\right)
+\frac{s(M_W^2+M_Z^2)}{c_W^2}\frac{L_dL_u}{tu}\cr
&&-\frac{s_W}{c_W}(\delta_Z W_{\delta} +x_Z W_x +y_Z W_y)
\quad +\frac{1}{2}\left(\frac{ss_W}{s-M_W^2}\right)^2\bigg[ \delta_Z^2 A_0
+y_Z^2 A_{y^2}(M_Z^2) \cr
&&+x_Z^2 A_{x^2}(M_Z^2)
+2\delta_Zx_ZA_x+2\delta_Zy_ZA_y +2x_Zy_ZA_{xy}(M_Z^2)\bigg]\cr
&&+\frac{s_W^2}{2}\frac{s}{s-M_W^2}\frac{s\beta^2}{M_W^2}z_Z
\cdot\Bigg[ \eta\Bigg(
\frac{s^3}{(s-M_W^2)tu2s_Wc_W}\bigg\{4\beta\cos\theta
c_W^2\frac{M_Z^2}{s}\cr
&&-\frac{1}{3}s_W^2\sin^2\theta
\bigg[1-\frac{M_W^2-2M_Z^2}{s}+\frac{M_Z^4-2M_W^2M_Z^2-M_W^4}{s^2}
-\frac{M_W^2(M_Z^4-M_W^4)}{s^3}\bigg]\cr
&&-\beta\cos\theta\sin^2\theta c_W^2\bigg[\frac{M_W^2+M_Z^2}{M_W^2}
+\frac{M_Z^2}{s}\frac{(2M_W^2-M_Z^2)}{M_W^2}
+\frac{M_W^2(M_Z^2-M_W^2)}{s^2}\bigg]\bigg\}\cr
&&+\frac{s\beta\cos\theta}{s-M_W^2}\frac{s}{M_W^2}(2\delta_Z+x_Z+y_Z)
+\frac{2s_WsM_Z^2}{3c_Wtu}\Bigg)\cr
&&+\frac{s}{s-M_W^2}\frac{s\beta^2}{4M_W^2}\Big\{
\Big(\frac{s}{M_W^2}-2\Big)(1+\cos^2\theta)+4\Big\}z_Z\Bigg]\cr
&&+\frac{s_W^2}{8}\left(\frac{s}{s-M_W^2}\right)^2\Bigg[
(z'_{3Z})^2\frac{4s^3\beta^4}{M_W^4M_Z^2}(1+\cos^2\theta)\cr
&&+(z'_{1Z})^2\frac{\beta^2}{M_W^2M_Z^2}\bigg\{\bigg[
(s-M_W^2-M_Z^2)^2+sM_Z^2-M_Z^4\bigg]\sin^2\theta+2sM_Z^2\bigg\}\cr
&&+(z'_{2Z})^2\frac{1}{M_W^2}\bigg\{s-2M_Z^2+\beta^2\cos^2\theta
(s-2M_W^2)+\frac{M_Z^4}{s}-3\frac{M_W^4}{s}\cr
&&+10\frac{M_W^2M_Z^2}{s}+2\frac{M_W^2}{s^2}(M_W^2-M_Z^2)^2\bigg\}\cr
&&+z'_{1Z}z'_{2Z}\eta\cos\theta\frac{4\beta}{M_W^2}(s-M_W^2-M_Z^2)
-z'_{2Z}z'_{3Z}\frac{8s}{M_W^2}\beta^2(1+\cos^2\theta)\Bigg]
\Bigg\},
\end{eqnarray}
with
\begin{equation}
t=\left\{ \begin{array}{c} (p_u-p_{W^+})^2 \\
                           (p_{\bar u}-p_{W^-})^2
        \end{array}
\right..
\end{equation}

The invariant functions for $q\bar{q}'\to W^{\pm}V_4$ for the
terms linear in the anomalous couplings are given by
\begin{eqnarray}
W_{\delta}&=&\frac{1}{(s-M_W^2)^2}\bigg[(tu-M_W^2M_Z^2)\left(\frac{s}{M_W^2}
c_W^2 -9c_W^2-1\right)\cr
&&\quad\quad\quad\quad\quad
\quad+2s(M_W^2+M_Z^2)\left(\frac{s}{M_W^2}c_W^2+3c_W^2-1\right)\bigg]\cr
&&+\frac{2}{s-M_W^2}\left(\frac{L_d}{t}-\frac{L_u}{u}\right)
\left(tu-M_W^2M_Z^2-s(M_W^2+M_Z^2)\right),\cr
W_x&=&\frac{1}{(s-M_W^2)^2}\bigg[s(s+3M_W^2-M_Z^2)-(tu-M_W^2M_Z^2)
(1+4c_W^2)\bigg]\cr
&&+\frac{1}{s-M_W^2}\left(\frac{L_d}{t}-\frac{L_u}{u}\right)
(tu-M_W^2M_Z^2-sM_Z^2),\cr
W_y&=&\frac{2s}{(s-M_W^2)^2}(s+3M_W^2-M_Z^2)-\frac{2s}{s-M_W^2}
\left(\frac{L_d}{t}-\frac{L_u}{u}\right)M_Z^2,
\end{eqnarray}
and the functions for the terms quadratic in the couplings are given by
\begin{eqnarray}
A_0&=&\left(\frac{tu}{M_W^2M_Z^2}-1\right)\left(\beta^2+
\frac{12M_W^2M_Z^2}{s^2}\right)
+\frac{2s(M_W^2+M_Z^2)}{M_W^2M_Z^2}\beta^2,\cr
A_x&=&\frac{s}{M_W^2}\beta^2-\frac{M_Z^2}{s}\left(\frac{tu}{M_W^2M_Z^2}-1
\right)\frac{(s-M_Z^2-5M_W^2)}{s},\cr
A_y&=&2\frac{s}{M_W^2}\beta^2,\cr
A_{x^2}(M_4^2)&=&\frac{s}{2M_W^2}\beta^2
-\frac{(tu-M_W^2M_4^2)}{sM_W^2}\frac{(s-2M_W^2-M_4^2)}{s},\cr
A_{xy}(M_4^2)&=&\frac{s}{2M_W^2}\beta^2 +\frac{(tu-M_W^2M_4^2)}{sM_W^2},\cr
A_{y^2}(M_4^2)&=&\frac{(tu-M_W^2M_4^2)}{M_W^4}\frac{(2s-M_W^2-M_4^2)}{s}
+\frac{s(M_W^2+M_4^2)}{2M_W^4}\beta^2.
\end{eqnarray}

\vspace{2mm}
\centerline{\underline{$q\bar{q}\to W^+W^-$}}
\nopagebreak

\begin{eqnarray}
\frac{d\sigma}{d\cos\theta}&=&\frac{\pi\alpha^2\beta}{24s_W^4s}\Bigg\{
\frac{(tu-M_W^4)}{s^2}\bigg[3-\frac{(s-6M_W^2)}{(s-M_Z^2)}\left(
\frac{L_q}{\tau_3^q}\right)\frac{1}{c_W^2}\cr
&&\quad~~~~~~~~\quad\quad\quad\quad
+\left(\frac{s}{s-M_Z^2}\right)^2\left(\beta^2+\frac{12M_W^4}{s^2}\right)
\left(\frac{L_q^2+R_q^2}{4c_W^2}\right)\bigg]\cr
&&-\frac{4M_Z^2}{s-M_Z^2}\left(\frac{L_q}{\tau_3^q}\right)
+\frac{s\beta^2M_Z^2}{(s-M_Z^2)^2}\frac{(L_q^2+R_q^2)}{c_W^2}\cr
&&+2\left(1+\frac{M_Z^2}{s-M_Z^2}\left(\frac{L_q}{\tau_3^q}\right)\right)
\left(\frac{tu-M_W^4}{st}-2\frac{M_W^2}{t}\right)
+\frac{tu-M_W^4}{t^2}\cr
&&-\frac{s_W}{c_W}(Z_{\delta}\delta_Z +Z_x x_Z +Z_y y_Z)
-s_W^2(\Gamma_x x_{\gamma} +\Gamma_y y_{\gamma})\cr
&&+\frac{1}{4}\frac{s_W^2}{c_W^2}\left(\frac{s}{s-M_Z^2}\right)^2
(L_q^2+R_q^2)\Big[ \delta_Z^2B_0 +x_Z^2 B_{x^2} +y_Z^2 B_{y^2}\cr
&&~~~~~~~~~~~~~~~~~~~~~~~~~~~~~~~~~\quad
+2\delta_Z x_Z B_x +2\delta_Z y_Z B_y +2x_Z y_Z B_{xy} \Big] \cr
&&+s_W^2\frac{s_W}{c_W}\frac{s}{s-M_Z^2}(L_q+R_q)Q_q\Big[ x_Zx_{\gamma}B_{x^2}
+y_Zy_{\gamma}B_{y^2} +\delta_Z x_{\gamma}B_x +\delta_Z y_{\gamma} B_y\cr
&&+(x_Z y_{\gamma} +x_{\gamma} y_Z) B_{xy}\Big]
+2s_W^4Q_q^2\Big[ x_{\gamma}^2 B_{x^2} +y_{\gamma}^2 B_{y^2}
+2x_{\gamma}y_{\gamma}B_{xy}\Big]\Bigg\},
\end{eqnarray}
with
\begin{equation}
t=\left\{ \begin{array}{c} (p_u-p_{W^+})^2 \\
                           (p_{\bar d}-p_{W^+})^2
        \end{array}
\right..
\end{equation}

The invariant functions for $q\bar{q}\to W^+W^-$ for the
terms linear in the anomalous couplings are given by
\begin{eqnarray}
Z_{\delta}&=&\frac{s}{s-M_Z^2}\bigg[\frac{M_W^2}{s}\left(\frac{tu}{M_W^4}-1
\right)
\left(1-6\frac{M_W^2}{s}-2\frac{M_W^2}{t}\right)
  +4\left(1+\frac{M_W^2}{t}\right)\bigg]
\left(\frac{L_q}{\tau_3^q}\right)\cr
&&-\frac{1}{2}\left(\frac{s}{s-M_Z^2}\right)^2
B_0\frac{M_Z^2}{s}(L_q^2+R_q^2),\cr
Z_x&=&\frac{s}{s-M_Z^2}\bigg[\frac{M_W^2}{s}\left(\frac{tu}{M_W^4}-1\right)
+2\left(1+\frac{M_W^2}{t}\right)\bigg]\left(\frac{L_q}{\tau_3^q}\right)
-\frac{1}{2}\left(\frac{s}{s-M_Z^2}\right)^2B_x\frac{M_Z^2}{s}(L_q^2+R_q^2),\cr
Z_y&=&2\left(1+\frac{M_W^2}{t}\right)\frac{s}{s-M_Z^2}
\left(\frac{L_q}{\tau_3^q}\right)
-\frac{1}{2}\left(\frac{s}{s-M_Z^2}\right)^2B_y\frac{M_Z^2}{s}(L_q^2+R_q^2),\cr
\Gamma_x&=&\frac{M_Z^2}{s-M_Z^2}|Q_q|\bigg[\left(\frac{tu}{M_W^4}-1\right)
(1-2s_W^2)
+\frac{s}{M_W^2}(2-4s_W^2)+4\bigg]\cr
&& +4s_W^2\frac{M_Z^2}{s-M_Z^2}B_xQ_q^2+4\frac{M_W^2}{t}|Q_q|,\cr
\Gamma_y&=&\frac{2}{c_W^2}\frac{s}{s-M_Z^2}|Q_q|\bigg[1-2s_W^2
+2\frac{M_W^2}{s}\bigg]
+4s_W^2\frac{M_Z^2}{s-M_Z^2}B_yQ_q^2 +4\frac{M_W^2}{t}|Q_q|,
\end{eqnarray}
and the functions for the terms quadratic in the couplings are given by
\begin{eqnarray}
B_0&=&\left(\frac{tu}{M_W^4}-1\right)\left(1-4\frac{M_W^2}{s}+12
\frac{M_W^4}{s^2}\right)
+4\frac{s}{M_W^2}\beta^2,\cr
B_x&=&\left(\frac{tu}{M_W^4}-1\right)\left(1-2\frac{M_W^2}{s}\right)
+2\frac{s}{M_W^2}\beta^2,\cr
B_y&=&2\frac{s}{M_W^2}\beta^2,\cr
B_{x^2}&=&\left(\frac{tu}{M_W^4}-1\right)\left(1-2\frac{M_W^2}{s}\right)
+\frac{s}{M_W^2}\beta^2,\cr
B_{xy}&=&-2\frac{M_W^2}{s}\left(\frac{tu}{M_W^4}-1\right)+\frac{s}{M_W^2}
\beta^2,\cr
B_{y^2}&=&2\left(\frac{tu}{M_W^4}-1\right)\left(1-\frac{M_W^2}{s}\right)
+\frac{s}{M_W^2}\beta^2.
\end{eqnarray}

\begin{table}
\hrule
\begin{center}
\begin{tabular}{|c|c|c|c|c|c|c|c|c||c|c|c|c|c|c|c|}\hline
\multicolumn{9}{|c||}{Leading Terms in $|{\cal M}|^2$, $q\bar{q}'\to
W^{\pm}Z$}&
\multicolumn{7}{|c|}{Leading Terms in $|{\cal M}|^2$, $q\bar{q}'\to
W^{\pm}\gamma$}\\ \hline\hline
$SM$&$\delta_Z$&$x_Z$&$y_Z$&$z_Z$&$\delta_Z^2$&$x_Z^2$&$y_Z^2$&$z_Z^2$&
$SM$&$x_{\gamma}$&$y_{\gamma}$&
$z_{\gamma}$&$x_{\gamma}^2$&$y_{\gamma}^2$&$z_{\gamma}^2$\\ \hline
$s^0$&$s$&$s^0$&$s^0$&$s\eta$&$s^2$&$s$&$s^2$&$s^3$&
$s^0$&$s^0$&/&$s\eta$&$s$&$s^2$&$s^3$\\ \hline\hline
\multicolumn{4}{|c|}{$\delta_Zx_Z,\delta_Zy_Z,x_Zy_Z$}&
\multicolumn{4}{|c|}{$z_Z\cdot(\delta_Z,x_Z,y_Z)$}&
\multicolumn{1}{|c||}{~}&
\multicolumn{2}{|c|}{$x_{\gamma}y_{\gamma}$}&
\multicolumn{3}{|c|}{$z_{\gamma}\cdot(x_{\gamma},y_{\gamma})$}&
\multicolumn{2}{|c|}{~}\\ \cline{1-8}\cline{10-14}
\multicolumn{4}{|c|}{$s$}&\multicolumn{4}{|c|}{$s^2\eta\cos\theta$}&
\multicolumn{1}{|c||}{~}&
\multicolumn{2}{|c|}{$s$}&\multicolumn{3}{|c|}{$s^2\eta\cos\theta$}&
\multicolumn{2}{|c|}{~}\\ \hline\hline
\multicolumn{2}{|c|}{$(z'_{1Z})^2$}&\multicolumn{2}{|c|}{$(z'_{2Z})^2$}&
\multicolumn{2}{|c|}{$(z'_{3Z})^2$}&\multicolumn{2}{|c|}
{$z'_{1Z}z'_{2Z}$}&\multicolumn{1}{|c||}{$z'_{2Z}z'_{3Z}$}&
\multicolumn{2}{|c|}{$(z'_{2\gamma})^2$}&\multicolumn{2}{|c|}
{$(z'_{3\gamma})^2$}&\multicolumn{3}{|c|}
{~}\\ \hline
\multicolumn{2}{|c|}{$s^2$}&\multicolumn{2}{|c|}{$s$}&
\multicolumn{2}{|c|}{$s^3$}&\multicolumn{2}{|c|}{$s\eta\cos\theta$}&
\multicolumn{1}{|c||}{$s$}&
\multicolumn{2}{|c|}{$s$}&\multicolumn{2}{|c|}{/}&
\multicolumn{3}{|c|}{~}\\ \hline
\end{tabular}
\end{center}
\caption{The leading behavior for $s\gg M_W^2$ of the helicity summed squared
amplitude $|{\cal M}|^2$ for $q\bar{q}'\to W^{\pm}Z$ and $q\bar{q}'\to
W^{\pm}\gamma$.
Shown is the leading behavior of the different terms in $|{\cal M}|^2$
proportional to the different combinations of the anomalous couplings.
If a term has a different sign for $W^+(\eta=-1)$ and $W^-(\eta=1)$ production
this is indicated by the factor $\eta$.
The terms can be even or odd in $\cos\theta$.
If they are odd this is indicated by the factor $\cos\theta$.
A slash indicates that a term is not present.
Sample usage: $|{\cal M}|^2= {O}(s^0) +\delta_Z {O}(s/M_W^2)
+\delta_Z^2 {O}(s^2/M_W^4)$ for $q\bar{q}'\to W^{\pm}Z$
if only $\delta_Z$ is non-zero. 
\label{hiamp}}
\hrule
\end{table}

Table \ref{hiamp} shows the behavior for $s\gg M_W^2$
for the helicity amplitudes for $q\bar{q}'\to W^{\pm}V$.
We note that the terms which are proportional to 
$\cos\theta$ give no contribution to
either the $p_T$ or the $M_{VV}$ distributions.

\section{Cross-Sections for \boldmath $W^{\pm}V_2\to W^{\pm}V_4$ 
\unboldmath Scattering\label{appb}}
\newcommand{\aw}{a_W}
\newcommand{\awp}{a_{W\Phi}}
\newcommand{\abp}{a_{B\Phi}}
We give expressions for cross-sections for $V_1V_2\to V_3V_4$
in a high-energy approximation (to be described below). 
We restrict ourselves to the
relevant processes of $WZ$ and $W\gamma$ production (in the following
we simply write $W$ instead of $W^\pm$). Thus, we 
only give the cross-sections for $WZ\to WZ, W\gamma\to WZ,
WZ\to W\gamma$ and $W\gamma\to W\gamma$.
Helicity amplitudes for processes $V_1V_2\to V_3V_4$
in the high-energy limit of the GIDS model can be found in
\cite{amps,dipl,uni}. They have been obtained from the exact Born-level 
amplitudes by an asymptotic expansion for $s\gg M_W^2$, where
$s$ is the scattering energy squared. The expansion
has been carried out at a fixed scattering angle $\theta$. Therefore,
also $|M_W^2/t|$ and $|M_W^2/u|$ have to be small parameters.
We note that for scattering energies $s>0.8$ TeV the parameters
$|M_W^2/t|$ and $|M_W^2/u|$ are smaller than 0.2 for all scattering
angles if a pseudorapidity cut of $\eta\le 1.5$ is applied.

Since we assume that the couplings are small, $\alpha_i={O}(M_W^2/
\Lambda^2)$, we only keep those anomalous
terms in which each power of an anomalous
coupling is enhanced by a factor of $s/4M_W^2$. 
For this purpose we define the parameters
\begin{equation}a_i\equiv \frac{s}{4M_W^2}\alpha_i.\end{equation}
The assumption $\alpha_i={O}(M_W^2/\Lambda^2)$ is equivalent to
assuming $a_i={O}(1)$ or smaller (since $s\le\Lambda^2$).
In addition to the non-standard terms, 
we include the leading standard terms, ${O}(s/4M_W^2)^0$.
We assume that the Higgs boson mass is small against the
scattering energy, $M_H^2\ll s$. 

Since the standard contributions to the amplitudes ${\cal M}_{++++}$ and
${\cal M}_{+--+}$ are very large (they diverge as $\cos\theta\to -1$), 
we also include the terms which arise when the sum of the leading standard
contribution and the subleading (i.e. first order
non-leading) non-standard contribution to these amplitudes is squared, i.e.
$2{\cal M}_{\mathrm{standard}}{\cal M}_{\mathrm{subleading}}$.
These terms are necessary to describe
the effects linear in the $\alpha_i$ for the cross-sections $\sigma_{TT}$
and $\sigma_{\overline{TT}}$ defined in (\ref{ttdef}).
The reason for this is that corresponding leading
terms are absent.

If the couplings are larger,
$a_i > 1$, also other subleading terms might
give sizeable contributions. Of the possible
non-standard subleading terms
those which are
quadrilinear in the couplings will be the largest ones. We include
also these terms. They only appear in amplitudes with an odd number
of longitudinally polarized vector bosons. 

The cross-sections for $pp\to WZ\to WZ$ and $pp\to W\gamma\to WZ$, 
calculated with the exact (numerically evaluated)
expressions for the cross sections for
$WZ\to WZ$ and $W\gamma\to WZ$, respectively, do not deviate by more
than 16\% from the same cross-sections calculated with the high-energy
approximation for the $WV\to WZ$ cross-sections presented here
if the $WV$ invariant mass is
in the range $0.8$ TeV $\le\sqrt{s}\le 2$ TeV.
This is true for couplings in the range $|\alpha_i|<0.1$.
This result was obtained for a pseudorapidity cut of $\eta=1$.
For $W\gamma$ production a similar result can be expected.

We give expressions for the integrated cross-sections summed
over the helicities of the outgoing particles. We write the cross-sections
as
\begin{eqnarray}
\sigma_{TT}&\equiv&\frac{1}{2}(\sigma_{++}+\sigma_{+-})\cr
\sigma_{\overline{TT}}&\equiv&\frac{1}{2}(\sigma_{++}-\sigma_{+-}),
\label{ttdef}\end{eqnarray}
with
\begin{eqnarray}
\sigma_{++}&=&\frac{C}{32\pi s}\frac{q}{p}\left(G_{++}^T
+G_{++++,\mathrm{subleading}}+G_{++00}\right.\cr
&&\left.\hspace{8mm}+G_{+++0}+G_{++-0}+G_{++0+}+G_{++0-}\right)\cr
\sigma_{+-}&=&\frac{C}{32\pi s}\frac{q}{p}\left(G_{+-}^T
+G_{+--+,\mathrm{subleading}}+G_{+-00}\right.\cr
&&\left.\hspace{8mm}+G_{+-+0}+G_{+--0}+G_{+-0+}+G_{+-0-}\right),
\end{eqnarray}
and
\begin{eqnarray}
\sigma_{TL}&=&\frac{C}{32\pi s}\frac{q}{p}\left(G_{+00+}+G_{+00-}+G_{+0+0}
+G_{+0-0}\right.\cr
&&\left.\hspace{8mm}+G_{+0++}+G_{+0--}+G_{+0+-}+G_{+0-+}+G_{+000}\right)\cr
\sigma_{LT}&=&\frac{C}{32\pi s}\frac{q}{p}\left(G_{0++0}+G_{0+-0}+G_{0+0+}
+G_{0+0-}\right.\cr
&&\left.\hspace{8mm}+G_{0+++}+G_{0+--}+G_{0++-}+G_{0+-+}+G_{0+00}\right)\cr
\sigma_{LL}&=&\frac{C}{32\pi s}\frac{q}{p}\left(G_{0000}+2G_{00++}+2G_{00+-}
+2G_{000+}+2G_{00+0}\right).
\end{eqnarray}
The quantities $G_{h_1h_2h_3h_4}$ are
the squared helicity amplitudes integrated over the scattering 
angle $\cos\theta$,
\begin{equation}
G_{h_1h_2h_3h_4}\equiv\frac{1}{C}\int\limits_{-z_0}^{z_0}\left|
{\cal M}_{h_1h_2h_3h_4}(\cos\theta)\right|^2 d\cos\theta,
\end{equation}
$C$ is a coupling factor which is different for each process 
and $z_0$ is an integration
limit for $|\cos\theta|$ determined e.g. by a cut.
The indices $h_1h_2h_3h_4$ 
denote the helicities of the particles $WV_2\to WV_4$ 
(in this order) and ${\cal M}_{h_1h_2h_3h_4}$ is the scattering-amplitude.
We further defined the sums over the transverse helicity amplitudes,
\begin{eqnarray}
G_{++}^T&\equiv&G_{++++,\mathrm{leading}}+G_{++--}+G_{+++-}+G_{++-+},\cr
G_{+-}^T&\equiv&G_{+--+,\mathrm{leading}}+G_{+-+-}+G_{+-++}+G_{+---}.
\end{eqnarray}
The expressions for $G_{++}^T$ and $G_{+-}^T$ are the same for all
processes $WV_2\to WV_4$, $V_i=\gamma,Z$,
\begin{eqnarray}
G_{++}^T&=&2c_W^4\Big[16z_0f_1
-8\aw^2(2\ln_1-7z_0-z_0^3)
-8\aw^3(3z_0-\frac{7}{3}z_0^3)\cr
&&\hspace{7mm}+\aw^4(9z_0+\frac{46}{3}z_0^3+\frac{z_0^5}{5})\Big],\cr
G_{+-}^T&=&2c_W^4\Big[16z_0f_1-16\ln_1+18z_0+\frac{2}{3}z_0^3
+2\aw^2(7z_0+\frac{5}{3}z_0^3)\cr
&&\hspace{7mm}+\aw^4(9z_0-\frac{2}{3}z_0^3+\frac{z_0^5}{5})\Big].
\label{tsums}\end{eqnarray}

$p$ and $q$ are the magnitudes of the three-momenta
of the vector-bosons in the initial and in the final state, respectively,
evaluated in the center-of-mass system of the vector-bosons,
given by
\begin{eqnarray}
p,q=\frac{\sqrt{s}}{2}\sqrt{1-\frac{2}{s}(M_W^2+
M_{i}^2)+\frac{1}{s^2}
(M_W^2-M_{i}^2)^2}\;\;,
\end{eqnarray}
where $i=2$ for $p$ and $i=4$ for $q$.
In (\ref{tsums}) and below we use the abbreviations
\begin{equation}
f_1\equiv\frac{1}{1-z_0^2},\hspace{7mm}
\ln_1\equiv\ln\left(\frac{1+z_0}{1-z_0}\right),\hspace{7mm}
r_H\equiv\frac{M_H^2}{M_W^2},
\end{equation}
and $t_W\equiv s_W/c_W$.
The coupling factors $C$ are given by
\begin{eqnarray}
C&=&g^4\;\,\mathrm{for}\;\, WZ\to WZ, \;\; C=g^4t_W^2\;\,\mathrm{for}\;\,
W\gamma\to WZ\;\,\mathrm{and}\;\, WZ\to W\gamma,\cr
C&=&g^4t_W^4\;\,\mathrm{for}\;\, W\gamma\to W\gamma.
\end{eqnarray}

\subsection{\boldmath $WZ\to WZ$\unboldmath }
For the process $WZ\to WZ$ there are 25 different 
helicity amplitudes ${\cal M}_{h_1h_2h_3h_4}$,
which cannot be related to each other by discrete symmetries ($C$, $P$
or $T$).
Of these amplitudes, 15 have leading terms of the order
${O}(s/4M_W^2)^0$, ${O}(a_i)$ or ${O}(a_i a_j)$.
Of these 15 terms, 6 only appear in the sums
$G_{++}^T$ and $G_{+-}^T$. 
The remaining 9 integrated squared amplitudes are given by
\begin{eqnarray}
G_{++00 \atop 00++}&=&2(\awp+\abp)^2s_W^2t_W^2z_0
+2c_W^2\aw^2(1-4\awp)^2\frac{z_0^3}{3}\cr
G_{+-00\atop 00+-}&=&\frac{s_W^2t_W^2}{2}\left(z_0+\frac{z_0^3}{3t_W^4}
\right)\cr
G_{0+0+}&=&\frac{c_W^4}{2}(1-t_W^2)^4z_0\cr
G_{0+0-}&=&2(1-2s_W^2)^2(\awp-t_W^2\abp)^2(z_0+\frac{z_0^3}{3})\cr
G_{+00+\atop 0++0}&=&8c_W^2z_0f_1 -2(1-2s_W^2)\ln_1 +\frac{z_0}{2c_W^2}
(1-2s_W^2)^2\cr
G_{+00-\atop 0+-0}&=&\hspace{3mm}\frac{c_W^2}{2}z_0\left((\awp+\abp)t_W^2
-3\aw-6\awp\aw\right)^2\cr
&&+\frac{z_0^3}{6}c_W^2\bigg[\left((\awp+\abp)t_W^2+\aw-4\awp\aw\right)^2\cr
&&\hspace{9mm}+4\left((\awp+\abp)t_W^2-3\aw-6\awp\aw\right)\awp\aw\bigg]\cr
&&+\frac{2}{5}z_0^5\awp^2\aw^2c_W^2\cr
G_{+0+0}&=&\frac{1}{2}z_0\cr
G_{+0-0}&=&2\awp^2(z_0+\frac{z_0^3}{3})\cr
G_{0000}&=&2z_0f_1-\frac{1}{2}\ln_1(2-r_H)+\frac{1}{8}z_0(9-2r_H+r_H^2)
+\frac{1}{24}z_0^3\cr
&&+\awp(12\ln_1-15z_0+3r_Hz_0-z_0^3)\cr
&&+\awp^2(-8\ln_1+43z_0+3r_Hz_0+\frac{z_0^3}{3}(25-r_H))\cr
&&+\awp^3(36z_0-28z_0^3)\cr
&&+\awp^4(18z_0+20z_0^3+2\frac{z_0^5}{5}).
\end{eqnarray}
The subleading terms for $G_{++++}$ and $G_{+--+}$ are given by
\begin{eqnarray}
G_{++++,\mathrm{subleading}}
&=&8c_W^2\mu_W\left[
\awp\left(8z_0f_1(2-s_W^2)+t_W^2(1-2s_W^2)\ln_1\right)\right.\cr
&&\hspace{12mm}\left.+\abp\left(-8z_0f_1s_W^2+t_W^2(1-2s_W^2)\ln_1\right)
\right.\cr
&&\hspace{12mm}\left.+\aw\left((5-2s_W^2)\ln_1-2z_0(3-s_W^2)\right)\right.\cr
&&\hspace{12mm}\left.+s_W^2t_W^2(\awp+\abp)^2\ln_1\right.\cr
&&\hspace{12mm}\left.+\aw(\ln_1-2z_0)
\left(\aw(3-s_W^2)-2\awp(2-s_W^2)+2\abp s_W^2\right)\right]\cr
G_{+--+,\mathrm{subleading}}
&=&2c_W^2\mu_W\left\{2(16z_0f_1-16\ln_1+17z_0+\frac{z_0^3}{3})\cdot
[\awp(2-s_W^2)-\abp s_W^2]\right.\cr
&&\left.\hspace{16mm}-(8\ln_1-14z_0-\frac{2}{3}z_0^3)\right.\cr
&&\left.\cdot\left[2\awp(\awp+t_W^2\abp)+\frac{1}{2}\awp(1+\frac{1}{c_W^2})
+\frac{1}{2}t_W^2\abp\right]\right\},
\end{eqnarray}
where we introduced the variable
\begin{equation}
\mu_W\equiv\frac{4M_W^2}{s}.
\end{equation}

The subleading terms 
which are quadrilinear in the couplings are given by,
\begin{eqnarray}
G_{+++0\atop +0++}&=&4\mu_Wc_W^2\aw^2(2\awp+\aw)^2(z_0-\frac{z_0^3}{3})\cr
G_{++-0\atop +0--}&=&\mu_Wc_W^2\aw^2(\awp+\aw)^2(9z_0-\frac{8}{3}z_0^3
-\frac{z_0^5}{5})\cr
G_{++0+\atop 0+++}&=&4\mu_W\aw^2\left(c_W^2(\awp-t_W^2\abp)+\awp
+c_W^2\aw\right)^2
\cdot(z_0-\frac{z_0^3}{3})\cr
G_{++0-\atop 0+--}&=&\mu_W\aw^2(\awp+c_W^2\aw)^2(9z_0-\frac{8}{3}z_0^3
-\frac{z_0^5}{5})\cr
G_{+-+0\atop +0+-}&=&\mu_Wc_W^2\aw^2\Big[(\awp^2+\aw^2)(z_0-\frac{z_0^5}{5})
\cr
&&\hspace{15mm}-2\awp\aw(z_0-\frac{2}{3}z_0^3+\frac{z_0^5}{5})\Big]\cr
G_{+--0\atop +0-+}&=&0\cr
G_{+-0+\atop 0++-}&=&0\cr
G_{+-0-\atop 0+-+}&=&\mu_W\aw^2\Big[4c_W^4(\awp-t_W^2\abp)^2(z_0
-\frac{z_0^3}{3})\cr
&&\hspace{19mm}+(\awp+c_W^2\aw)^2(z_0-\frac{z_0^5}{5})\cr
&&\hspace{11mm}-4c_W^2(\awp-t_W^2\abp)(\awp+c_W^2\aw)(z_0-\frac{z_0^3}{3})
\Big]\cr
G_{+000\atop 00+0}&=&\mu_W\awp^2(\aw+\awp)^2(9z_0-\frac{8}{3}z_0^3
-\frac{z_0^5}{5})\cr
G_{0+00\atop 000+}&=&\mu_W\frac{1}{c_W^2}\awp^2(\awp+c_W^2\aw)^2(9z_0
-\frac{8}{3}z_0^3-\frac{z_0^5}{5}).
\end{eqnarray}

\mathversion{bold}
\subsection{$W\gamma\to WZ$}
\mathversion{normal}
For $W\gamma\to WZ$ the cross-sections $\sigma_{TL}$ and $\sigma_{LL}$
vanish. There are 27 different amplitudes, out of which
14 have terms of the leading order.
Of these amplitudes, 8 only appear in the sums of
the transverse amplitudes $G_{++}^T$ and
$G_{+-}^T$.
For the remaining 6 helicity combinations, 
the integrated squared amplitudes are
given by the expressions,
\begin{eqnarray}
G_{++00}&=&2c_W^2\Big[\aw^2(1-4\awp^2)\frac{z_0^3}{3}
+(\awp+\abp)^2z_0\Big]\cr
G_{+-00}&=&\frac{c_W^2}{2}(z_0+\frac{z_0^3}{3})\cr
G_{0++0}&=&2c_W^2\left(4z_0f_1-2\ln_1+z_0\right)\cr
G_{0+-0}&=&\frac{c_W^2}{2}\left[(\awp+\abp)^2(z_0+\frac{z_0^3}{3})
+\aw^2(9z_0+\frac{z_0^3}{3})\right.\cr
&&\hspace{6mm}\left.+(\awp+\abp)\aw(6z_0-\frac{2}{3}z_0^3)\right.\cr
&&\hspace{6mm}\left.+2\awp\aw(\awp+\abp)(6z_0+\frac{2}{3}z_0^3)\right.\cr
&&\hspace{6mm}\left.+2\awp\aw^2(18z_0-\frac{10}{3}z_0^3)\right.\cr
&&\hspace{6mm}\left.+2\awp^2\aw^2(18z_0-\frac{4}{3}z_0^3+\frac{2}{5}z_0^5)
\right]\cr
G_{0+0+}&=&2(1-2s_W^2)^2z_0\cr
G_{0+0-}&=&2\left[\awp(\frac{3}{2}-2s_W^2)+\abp(\frac{1}{2}-2s_W^2)
\right]^2(z_0+\frac{z_0^3}{3}).
\end{eqnarray}
The non-leading terms for $G_{++++}$ und $G_{+--+}$ have the following
expressions,
\begin{eqnarray}
&&G_{++++,\mathrm{subleading}}\cr
&=&8c_W^2\mu_W\bigg[\awp\left(4z_0f_1(3-2s_W^2)
-\ln_1(\frac{1}{2}-2s_W^2)\right)\cr
&&\hspace{13mm}+\abp\left(4z_0f_1(1-2s_W^2)-\ln_1(\frac{1}{2}-2s_W^2)
\right)\cr
&&\hspace{13mm}+\aw\left(\ln_1(\frac{7}{2}-2s_W^2)-2z_0(2-s_W^2)\right)\cr
&&\hspace{13mm}-s_W^2\ln_1(\awp+\abp)^2\cr
&&\hspace{13mm}+(\ln_1-2z_0)\aw\left(\aw(2-s_W^2)
-\awp(3-2s_W^2)-\abp(1-2s_W^2)\right)\bigg]\cr
&&G_{+--+,\mathrm{subleading}}\cr
&=&2c_W^2\mu_W\Bigg\{\awp\bigg[16z_0f_1(3-2s_W^2)-\ln_1(50-32s_W^2)\cr
&&\hspace{25mm}+\frac{z_0}{2}(109-68s_W^2)+\frac{z_0^3}{3}(7-4s_W^2)\bigg]
\cr
&&\hspace{15mm}+\abp\bigg[32z_0f_1c_W^2-\ln_1(22-32s_W^2)\cr
&&\hspace{25mm}+\frac{z_0}{2}(35-68s_W^2)+\frac{z_0^3}{6}(1-4s_W^2)\bigg]
\cr
&&\hspace{15mm}
+\awp(\awp+\abp)(8\ln_1-14z_0-\frac{2}{3}z_0^3)\Bigg\}.
\end{eqnarray}
The subleading terms which are quadrilinear in
the couplings are given by,
\begin{eqnarray}
G_{+++0}&=&4c_W^2\mu_W\aw^2(2\awp-\aw)^2(z_0-\frac{z_0^3}{3})\cr
G_{++-0}&=&c_W^2\mu_W\aw^2(\awp+\aw)^2(9z_0-\frac{8}{3}z_0^3
-\frac{z_0^5}{5})\cr
G_{++0+}&=&4\mu_W\aw^2\left[\awp+c_W^2\aw
+c_W^2(\awp-t_W^2\abp)\right]^2(z_0-\frac{z_0^3}{3})\cr
G_{++0-}&=&\mu_W\aw^2\left[c_W^4\aw^2(9z_0-\frac{8}{3}z_0^3
-\frac{z_0^5}{5}\right.)\cr
&&\hspace{15mm}\left.+12c_W^2\aw(\awp-\abp)(z_0-\frac{z_0^3}{3})\right.\cr
&&\hspace{15mm}\left.+4(\awp-\abp)^2(z_0-\frac{z_0^3}{3})\right]\cr
G_{+-+0}&=&c_W^2\mu_W\aw^2\left[\aw^2(z_0-\frac{z_0^3}{3})
-2\aw\awp(z_0-\frac{z_0^3}{3})
+\awp^2(z_0-\frac{z_0^5}{5})\right]\cr
G_{+--0}&=&0\cr
G_{+-0+}&=&0\cr
G_{+-0-}&=&c_W^4\mu_W\aw^2\left[\aw^2(z_0-\frac{z_0^5}{5})
-4\aw(\awp+\abp)(z_0-\frac{z_0^3}{3})\right.\cr
&&\hspace{17mm}\left.+4(\awp+\abp)^2(z_0-\frac{z_0^3}{3})\right]\cr
G_{0+++}&=&4c_W^4\mu_W\aw^2(\awp+\abp+\aw)^2(z_0-\frac{z_0^3}{3})\cr
G_{0+--}&=&\mu_W\aw^2\left[4(\abp+c_W^2\aw)^2(z_0-\frac{z_0^3}{3}
)\right.\cr
&&\hspace{15mm}
\left.+4(\awp+c_W^2\aw)(\abp+c_W^2\aw)(z_0-\frac{z_0^3}{3})\right]\cr
G_{0++-}&=&0\cr
G_{0+-+}&=&\mu_W\aw^2\left[(\awp+c_W^2\aw)^2(z_0-\frac{z_0^5}{5}
)\right.\cr
&&\hspace{14mm}\left.-4(\awp+c_W^2\aw)c_W^2(\awp-t_W^2\abp)(z_0-\frac{z_0^3}{3}
)\right.\cr
&&\hspace{14mm}
\left.+4c_W^2(\awp-t_W^2\abp)^2(z_0-\frac{z_0^3}{3})\right]\cr
G_{0+00}&=&\mu_Wc_W^2\awp^2\aw^2(9z_0-\frac{8}{3}z_0^3
-\frac{2}{5}z_0^5).
\end{eqnarray}

\mathversion{bold}
\subsection{$WZ\to W\gamma$}
\mathversion{normal}
The terms $G_{h_1h_2h_3h_4}$ for $WZ\to W\gamma$ can be obtained from the terms
$G_{h_1h_2h_3h_4}$ for $W\gamma\to WZ$ by exchanging the helicity indices 
according to
$G_{h_1h_2h_3h_4}^{WZ\to W\gamma}=G_{h_3h_4h_1h_2}^{W\gamma\to WZ}$.
Also a parity transformation
$G^{WZ\to W\gamma}_{h_1h_2h_3h_4}=
G^{WZ\to W\gamma}_{-h_1-h_2-h_3-h_4}$ might have to be applied.
The combinations with $h_4=0$ vanish.

\mathversion{bold}
\subsection{$W\gamma\to W\gamma$}
\mathversion{normal}
For $W\gamma\to W\gamma$ the cross-sections $\sigma_{TL}$ and
$\sigma_{LL}$ and the terms $G_{h_1h_2h_3h_4}$ with
$h_4=0$ vanish. There are 12 different helicity combinations
which can not be related to each other by discrete symmetries.
8 combinations receive leading contributions.
6 of the latter combinations only enter the expressions $G_{++}^T$ and
$G_{+-}^T$. The remaining two leading terms are given by
\begin{eqnarray}
G_{0+0+}&=&8c_W^4z_0\cr
G_{0+0-}&=&8c_W^4(\awp+\abp)^2(z_0+\frac{1}{3}z_0^3).
\end{eqnarray}
The subleading terms for $G_{++++}$ and $G_{+--+}$ are given by
\begin{eqnarray}
G_{++++,\mathrm{subleading}}&=&\mu_Wc_W^4
\Big[ 16(\awp+\abp)(4z_0f_1-\ln_1)
+16\aw(\ln_1-z_0)\cr
&&+16\aw(\awp+\abp)(2z_0-\ln_1) -8\aw^2(2z_0-\ln_1)+8(\awp+\abp)^2\ln_1
\Big]\cr
G_{+--+,\mathrm{subleading}}&=&4c_W^4\mu_W(\awp+\abp)\Big[ 16(z_0f_1-\ln_1)
+17z_0+\frac{1}{3}z_0^3\Big].
\end{eqnarray} 
The subleading terms
which are quadrilinear in the couplings are given by,
\begin{eqnarray}
G_{++0+\atop 0+++}&=&4\mu_Wc_W^4
\aw^2(\awp+\abp+\aw)^2(z_0-\frac{1}{3}z_0^3)\cr
G_{++0-\atop 0+--}&=&\mu_Wc_W^4\aw^4(9z_0-\frac{8}{3}z_0^3-\frac{1}{5}z_0^5)\cr
G_{+-0+\atop 0++-}&=&0\cr
G_{+-0-\atop 0+-+}&=&\mu_Wc_W^4\Big[\aw^4(z_0-\frac{1}{5}z_0^5)
-4\aw^3(\awp+\abp)(z_0-\frac{1}{3}z_0^3)\cr
&&\hspace{7mm}+4\aw^2(\awp+\abp)^2(z_0-\frac{1}{3}z_0^3)\Big].
\end{eqnarray}

\end{appendix}

\end{document}